\documentclass[aps,amsmath,twocolumn,amssymb,pre,showpacs]{revtex4}
\usepackage{amsmath,amssymb}
\usepackage{graphicx}
\usepackage{epsfig}
\usepackage{color}

\bibstyle{apsrev.bib}
\newcommand{\be}{\begin{equation}}
\newcommand{\ee}{\end{equation}}
\newcommand{\beqn}{\begin{eqnarray}}
\newcommand{\eeqn}{\end{eqnarray}}

\begin{document}  
\input epsf.sty

\title{Distribution of dynamical quantities in the contact process, random walks, and quantum spin chains in random environments}

\author{R\'obert Juh\'asz}
\affiliation{Institute for Solid State Physics and Optics, Wigner Research Centre for Physics, H-1525 Budapest, P.O. Box 49, Hungary}
\email{juhasz.robert@wigner.mta.hu}
\date{\today}

\begin{abstract}
We study the distribution of dynamical quantities in various one-dimensional, disordered models the critical behavior of which is described by an infinite randomness fixed point.
In the {\it disordered contact process}, the survival probability 
$\mathcal{P}(t)$ is found to show multi-scaling in the critical point, meaning that $\mathcal{P}(t)=t^{-\delta}$, where the (environment and time-dependent) exponent $\delta$ has a universal limit distribution when $t\to\infty$. 
The limit distribution is determined by the strong disorder renormalization group method analytically in the end point of a semi-infinite lattice, where it is found to be exponential,  while, in the infinite system, conjectures on its limiting behaviors for small and large $\delta$, which are based on numerical results, are formulated. 
By the same method, the survival probability in the problem of {\it random walks in random environments} is also shown to exhibit multi-scaling with an exponential limit distribution. 
In addition to this, the (imaginary-time) spin-spin autocorrelation function of the {\it random transverse-field Ising chain} is found to have a form similar to that of survival probability of the contact process at the level of the renormalization approach. Consequently, a relationship between the corresponding limit distributions in the two problems can be established. Finally, the distribution of the spontaneous magnetization in this model is also discussed. 
\end{abstract}

\pacs{02.50.Ey, 64.60.Ak, 87.23.Cc, 75.10.Nr}

\maketitle

\section{Introduction}

Spatial heterogeneity has a substantial effect on systems with many degrees of freedom, especially in the vicinity of phase transitions. 
Local quantities become, in general, non-self-averaging --- 
 their average and typical value is different, or even may obey to different scaling laws at criticality \cite{wd}. 
This phenomenon is most prominently present in
 strongly disordered models, where the critical and even the off-critical behavior is determined by the fluctuations of disorder \cite{im,vojta_rev}. 
Though perturbative treatments of disorder or various effective medium approximations fail to work in these models, by coarse-graining the system via appropriate decimation procedures, the strength of disorder is growing without limits, and this is the property that makes these problems tractable \cite{im}. 
The critical behavior of these models is controlled by an {\it infinite randomness fixed point} (IRFP), at which many asymptotic properties have been calculated analytically in one dimension and numerically in higher dimensions by means of the aforementioned decimation scheme, also known as {\it strong disorder renormalization group} (SDRG) method.  
Examples of this class of models are disordered quantum spin chains at zero temperature \cite{fisher}, random walks in random environments \cite{sk,fdm}, or the disordered contact process \cite{noest,hiv}. 
In this work, we will deal with some common dynamical properties of these models.  Although these features are formally similar, 
their particular appearance, such as which quantity carries that property, may be model-dependent.  
The results will be expounded first in the language of the contact process, as we think that this is the most expressive way of presentation. 
Thereafter, similar features of random walks and quantum spin chains will be enlightened. 
 
The {\it contact process} \cite{cp,liggett} is a stochastic model consisting of two competing processes ---  reproduction and death of individuals living on a lattice, and has applications mainly in population dynamics and in the modeling of epidemic spreading.  
Tuning the relative rates of the above processes, the model undergoes an absorbing phase transition from a fluctuating active phase with a finite density of individuals to an inactive one, falling in the directed percolation universality class \cite{md,odor,hhl}.  

In the spatially inhomogeneous model, where the reproduction and death rates are site-dependent, random variables, the SDRG transformation \cite{hiv} is essentially identical to that of the random transverse-field Ising model \cite{fisher}, and drives the critical model to an IRFP, at least for strong enough disorder of the parameters. The critical behavior of the latter model for an arbitrary weak disorder is described by the IRFP in one-dimension but, whether the same is true for the contact process is not yet known, as this question cannot be decided leaning solely on the SDRG method. 
Nevertheless, large scale Monte Carlo simulations support the predictions of the method in $d=1,2,3$ dimensions even for relatively weak disorder strengths \cite{vojta}. 

The information on time-dependent quantities gained by the SDRG procedure has been so far restricted to the {\it averages} taken over random environments (i.e. set of transition rates). 
This is founded on the assumption that, as the local densities depend strongly 
on the particular realization of the underlying random environment, 
the average is dominated by certain atypical environments that are locally supercritical on a given time-scale and whose fraction is vanishing with the time. 
An example is the survival probability of the contact process, which is defined as the probability that, starting with a single individual, the absorbing state has not yet reached up to time $t$. 
Its average over random environments decays ultra-slowly, as an inverse power of the logarithm of time in the critical point due to the contribution of atypical realizations \cite{moreira,hiv}. But the survival probability in a typical realization of the environment is expected to decay much more rapidly. 
In applications, for instance, in case of a population placed in a static random environment (or at least, in an environment changing on a much longer time scale than the population) the relevant quantity characterizing the fate of the population is the survival probability in that environment as opposed to an average over a set of other ``possible'' environments. 
In spite of this, the investigation of the average behavior of dynamical quantities was in focus so far and, apart from certain autocorrelation functions of the random transverse-field Ising chain \cite{ky,ijr}, no attention has been paid to the typical behavior. 

In this work, we wish to make steps in this direction, aiming at studying the distribution of time-dependent local quantities in one dimensional models characterized by an IRFP in the critical point.  
To this end, we will apply the SDRG method, which has been so far used to 
infer the dynamical scaling of the averages in an indirect way, 
for calculating dynamical quantities in individual realizations of the random environment. 
In particular, we consider the survival probability in the contact process and in random walks, furthermore, in the random transverse-field Ising chain,  
the distribution of the spontaneous magnetization (which has been studied in a semi-infinite chain in Ref. \cite{fisher}) and 
the distribution of the imaginary-time spin-spin autocorrelation function. 
Calculations in the end point of a semi-infinite chain are carried out analytically, while, for bulk quantities, mainly numerical results are in our disposal.
In case of the contact process, the results obtained by the SDRG are also compared with those of Monte Carlo simulations. 
The above quantities are found to exhibit multi-scaling \cite{ky}, meaning they are an inverse power of the time but, instead of a single exponent 
characteristic for homogeneous systems, a broad distribution of random, environment-dependent exponents appears here. 

The outline of the paper is as follows. 
In section \ref{model}, the contact process is defined and the SDRG approach is reviewed. In section \ref{scaling}, the behavior of the average survival probability is recalled and a phenomenological scaling theory of the distribution of the survival probability is presented. 
The end point survival probability is calculated by the SDRG approach and compared to results of Monte Carlo simulations in section \ref{endpoint}, while,
 in section \ref{bulk},  
the way of numerical calculation of the bulk survival probability by the SDRG 
method is presented together with numerical results. 
In section \ref{rwre}, the scaling function of the distribution of the survival probability in the problem of random walks in random environments is calculated in the frame of the SDRG method. 
The distribution of the spontaneous magnetization and imaginary-time spin-spin autocorrelation function of the random transverse-field Ising chain is discussed in section \ref{quantum}. 
Finally, the results are summarized and an outlook on further work is given in section \ref{summary}.

\section{Renormalization approach to the contact process}
\label{model}

Let us consider a one-dimensional lattice the sites of which are labeled by the integers. On each site, a binary variable is defined, which encodes that the site is (using the terminology of population dynamics) empty or occupied by an individual. The contact process is a continuous-time Markov process with the following independent transitions. An individual on site $i$ produces an offspring on site $i+1$ with rate $\nu_i$ and on site $i-1$ with rate $\kappa_{i-1}$. Furthermore, an individual on site $i$ dies with a rate $\mu_i$. The reproduction and death rates are independent, identically distributed random variables, for which we impose that the distributions of rates are invariant under the interchange of $\nu_i$ and $\kappa_{i-1}$ so that the model is left-right symmetric in a statistical sense.  

An SDRG approach was formulated first to the symmetric variant of this model (where $\nu_i=\kappa_{i-1}$) in Ref. \cite{hiv} and it has later been generalized to the non-symmetric case \cite{juhasz}. 
The SDRG is a real-space decimation procedure, which sequentially reduces the degrees of freedom in small blocks of sites, dropping thereby the high-lying levels of the Liouville operator $Q$ that governs the evolution of the state 
$|P(t)\rangle$ through the master equation 
\be 
\partial_t|P(t)\rangle=-Q|P(t)\rangle.
\label{master}
\ee
The low-lying part of the spectrum of $Q$, which is responsible for the long-time dynamics, is conserved by the procedure, asymptotically exactly. 
The decimation procedure consists of two kinds of reduction steps. 
First, if $\lambda_1\gg \mu_1,\mu_2$ in a block of two sites, where $\lambda_i\equiv\frac{\nu_i\kappa_i}{\nu_i+\kappa_i}$ is a combined reproduction rate, it is replaced by a single site with an approximate death rate obtained perturbatively as 
\be 
\tilde\mu\simeq\frac{\mu_1\mu_2}{\lambda_1}.
\label{mu_appr}
\ee
Second, if the death rate at site $2$ is large, such that $\mu_2\gg\lambda_1,\lambda_2$, it is eliminated and site $1$ and $3$ are directly connected by a link with approximate reproduction rates
\be 
\tilde\nu\simeq\frac{\nu_1\nu_2}{\mu_2},
\qquad 
\tilde\kappa\simeq\frac{\kappa_1\kappa_2}{\mu_2}.  
\label{nu_appr}
\ee
The precise forms of the effective rates can be found in Ref. \cite{juhasz}. 
Close to the fixed point of the transformation, the asymmetry between 
$\nu_i$ and $\kappa_{i-1}$ becomes irrelevant and, instead of these variables, it is sufficient to deal with the combined reproduction rate, which transforms as
\be
\tilde\lambda\simeq 2\frac{\lambda_1\lambda_2}{\mu_2}.
\label{lambda_appr}
\ee
The decimation procedure consists of iteratively finding the largest rate $\Omega$ in the set of transition rates $\{\lambda_i,\mu_i\}$ and applying the first and second reduction step if $\Omega=\lambda_i$  and $\Omega=\mu_i$, 
respectively. 
Under the iterations, the rate scale $\Omega$ gradually decreases starting from some initial value $\Omega_0$, while 
the length scale $\xi$ (the ratio of the number of sites in the original model to that of the renormalized one) increases. Due to the first kind of step, the sites of the renormalized model are, in fact, clusters composed of many original (not necessarily adjacent) lattice sites and their mass $m$, i.e. the number of original sites in them increases under the renormalization. 
In the critical model, the distribution of logarithmic rates, which remain independent under the renormalization, are broadening without limits, which is characteristic for an IRFP. 
The fixed point solution of the renormalization flow equations formulated 
for the probability densities of rates are pure exponentials 
\be 
\tilde P(\theta)=e^{-\theta}, \quad  \tilde R(\eta)=e^{-\eta}.
\label{fpsol} 
\ee
in terms of the rescaled rates $\theta=p_0(\Gamma)\beta$, $\eta=r_0(\Gamma)\zeta$, where 
\be
\beta=\ln(\Omega/\mu)  \qquad \zeta=\ln(\Omega/\lambda)
\ee 
are logarithmic variables and the scale factors 
$p_0(\Gamma)$ and $r_0(\Gamma)$ are
\be
p_0(\Gamma)=r_0(\Gamma)=\Gamma^{-1}\equiv[\ln(\Omega_0/\Omega)]^{-1}. 
\label{pnullcp}
\ee
in the critical point \cite{fisher}. 
The limit distributions of rates $\nu$ and $\kappa$ are identical to that of $\lambda$. 
Furthermore, the rate scale $\Omega$ and the average mass $m$ varies with the 
renormalized length scale $\xi$ as 
\be 
\ln(\Omega_0/\Omega)\sim \xi^{\psi}  \qquad 
m\sim \xi^{1-x_b}
\label{omegaxicp}
\ee
with the critical exponents $\psi=1/2$ and $x_b=(3-\sqrt{5})/4$.  

Below the critical point, in the absorbing phase, the renormalization flow is 
attracted by a line of fixed points parametrized by a disorder-dependent 
dynamical exponent $z$, which diverges as the critical point is approached.
Here, the fixed point distributions of rates are of the same form as given in Eqs. (\ref{fpsol}) but the scale factors are \cite{fisher,igloi}  
\be
p_0(\Omega)-1/z=r_0(\Omega)=O(\Omega^{1/z}), 
\label{pnullgp}
\ee 
and the basic relationships for $\Omega$ and $m$ read as  
\be 
\Omega\sim \xi^z  \qquad m\sim \ln\xi. 
\label{omegaxigp}
\ee
This phase of the contact process \cite{noest} is analogous to the Griffiths-McCoy phase of random quantum spin models \cite{griffiths}.

\section{Scaling theory of the survival probability}
\label{scaling}

\subsection{Quantities of interest}

The time-dependent survival probability is a frequently studied dynamical quantity in the contact process. In a given realization 
$\{\nu_n,\kappa_{n-1},\mu_n\}$ of the random environment it is defined as the probability that there is at least one individual on the lattice at time $t$ provided the process was started with a single individual on site $i$. 
It will be denoted by $\mathcal{P}_i(t;\{\nu_n,\kappa_{n-1},\mu_n\})$ or, briefly by $\mathcal{P}(t)$. 
The average $\overline{\mathcal{P}(t)}$ of the survival probability over random environments (or, equivalently, over starting sites in an individual realization of the environment), taken in the limit $t\to\infty$, is the order parameter of the absorbing phase transition. 

In terms of the left and right eigenstates $\langle n|$ and $|n\rangle$, respectively, of the Liouville operator pertaining to the eigenvalue $\epsilon_n$, 
$\mathcal{P}(t)$ assumes the form
\be 
\mathcal{P}(t)=-\sum_{n>0}\langle n|P(0)\rangle\langle 0|n\rangle e^{-\epsilon_nt}. 
\label{Pexact}
\ee
Here, $|P(0)\rangle$ and $|0\rangle$ denote the initial and the stationary (empty lattice) state, respectively, and the summation goes over the non-stationary states, for which $\epsilon_n>0$.

A closely related quantity is time-dependent local density $\rho_i(t;\{\nu_n,\kappa_{n-1},\mu_n\})$, which is the probability of finding an individual on site $i$ at time $t$, provided that, at time $t=0$, the lattice was completely occupied. 
The self-dual property of the symmetric model ($\nu_n=\kappa_{n-1}$) implies an exact equality $\mathcal{P}_i(t;\{\nu_n,\kappa_{n-1},\mu_n\})=\rho_i(t;\{\nu_n,\kappa_{n-1},\mu_n\})$ for any realization $\{\nu_n,\kappa_{n-1},\mu_n\}$ of the random environment \cite{hv}. In the general case, a slightly weaker relationship $\mathcal{P}_i(t;\{\nu_n,\kappa_{n-1},\mu_n\})=\rho_i(t;\{\kappa_{n-1},\nu_n,\mu_n\})$ holds, 
where, in the r.h.s., the rates $\nu_n$ and $\kappa_{n-1}$ are interchanged compared to the l.h.s. \cite{ascpmc}. 
Nevertheless, due to the invariance of the distribution of rates under 
this operation, the probability distributions of the above quantities 
will be identical:
\be
{\rm Prob}[\mathcal{P}(t);t]\equiv {\rm Prob}[\rho(t);t].
\ee 
Thus, all the results formulated for the distribution of $\mathcal{P}(t)$ in the sequel is also valid for that of the local density. 

\subsection{Contribution of atypical environments to the average}

The renormalization scheme reviewed in the previous section can be used to construct a scaling theory of the average survival probability as follows. 
By applying the SDRG procedure, each cluster is decimated out 
at some (sufficiently small) rate scale $\Omega$, which gives its effective death rate $\tilde\mu=\Omega$. 
When the contact process starts with a single occupied site, 
the whole cluster containing this site will be activated with a probability $O(1)$.
The probability that the activity spreads to other clusters is, however, small, and, if one is interested in the leading order time-dependence of the {\it average} survival probability, this effect is believed to be negligible 
\cite{hiv}. 
According to this, the dominant contribution to the 
the average survival probability at time $t$ is given by the probability that the starting site is contained in a cluster that is active (non-decimated) at the scale $\Omega=t^{-1}$ since these clusters have a mean lifetime $\tilde\mu^{-1}$ at least $t$. 
This contribution to the average, which comes from the rare, atypical starting sites described above, will be denoted by $\overline{\mathcal{P}_{\rm atyp}(t)}$. 
The above probability on the renormalized length scale $\xi$ is given by
$\xi^{-1}m(\xi)$. 
Using Eq. (\ref{omegaxicp}), this results in 
\be 
\overline{\mathcal{P}_{\rm atyp}(t)}\sim (\ln t)^{-x_b/\psi},
\label{pavcp}
\ee
in the critical point. 
If the starting site is the first site of a semi-infinite lattice, the 
probability that this site is still not decimated at scale $\Omega$ 
is \cite{fisher}
\be
{\rm Prob(1st~site~is~not~decimated~until~\Omega)}\sim [\ln(\Omega_0/\Omega)]^{-1}.
\label{sp}
\ee
This leads to 
\be 
\overline{\mathcal{P}_{\rm 1,atyp}(t)}\sim (\ln t)^{-1},
\label{pavcps}
\ee
where the decay exponent is different from that of the bulk $x_b/\psi\approx 0.382$. 

Following the above arguments in the Griffiths phase and using Eq. (\ref{omegaxigp}), one obtains 
\be 
\overline{\mathcal{P}_{\rm atyp}(t)}\sim t^{-1/z}\ln t.
\label{pavgp}
\ee
In the case of the first site of a semi-infinite lattice, the probability that it is not decimated out down to scale $\Omega$ can be calculated following Ref. \cite{fisher} with the result
\be 
{\rm Prob(1st~site~is~not~decimated~until~\Omega)}\sim (\Omega/\Omega_0)^{1/z}.
\label{sps}
\ee
The contribution from atypical environments to the average in this case decays purely algebraically as 
\be 
\overline{\mathcal{P}_{\rm 1,atyp}(t)}\sim t^{-1/z}.
\label{pavgps}
\ee

\subsection{Distribution of the survival probability}
\label{typscaling}

We have seen that the average survival probability at time $t$ is thought to be dominated by the fraction of lattice sites that are parts of clusters whose lifetime is at least $t$. This fraction is vanishing with increasing $t$. 
But how does the survival probability $\mathcal{P}(t)$ behave for long times if the starting site is a typical one?
Then, we have to follow $\mathcal{P}(t)$ up to times that may be much longer than the lifetime of the primary cluster containing the starting site and 
the fact that the activity may spread to other clusters can no longer be disregarded. 
Although this occurs with small probabilities, the activity can get 
in this way to clusters having a much longer lifetime than the primary one, influencing thereby the long-time behavior considerably. 
The survival probability, on the level of a phenomenological description, can thus be written in the form 
\be
\mathcal{P}(t)\approx \sum_np_ne^{-\tilde\mu_nt}
\label{ptsum}
\ee
where $p_n$ denotes the probability that the activity reaches a cluster of death rate $\tilde\mu_n$ and the summation goes over all clusters identified by the SDRG method.  
Here, $n=n_1$ corresponds to the primary cluster, so that $p_{n_1}=O(1)$.
 
Clearly, the terms with $\mu_n>\mu_{n_1}$ or even $\mu_n\approx\mu_{n_1}$ are negligible in the above sum as the corresponding clusters are farther and farther from the primary one and $p_n$ decreases rapidly with the distance (faster than any power of the distance). 
One can easily convince oneself that beside the term $n=n_1$ it is sufficient to keep the term $n=n_2$ with the smallest access probability $p_{n_2}$ among those having a death rate smaller than $\tilde\mu_1$, then the term with the smallest access probability $p_{n_3}$ among those having a death rate smaller than $\tilde\mu_{n_2}$, and so on. All the other terms do not substantially influence the sum and can be dropped.  
The essential terms of Eq. (\ref{ptsum}) can be arranged in a form where the subsequent terms correspond to clusters lying typically farther and farther from the primary cluster and having smaller and smaller access probabilities and death rates.
Notice that this approximate expression of $\mathcal{P}(t)$ has the same structure as the exact one in Eq. (\ref{Pexact}), but contains only the relevant terms, which are much less in number.  

Now, we can obtain a rough picture on the behavior of $\mathcal{P}(t)$ by using the known asymptotic scaling properties of reproduction and death rates quoted in section \ref{model}. 
In the critical point, the scaling variables 
$\frac{\ln(\Omega/\lambda)}{\ln(\Omega_0/\Omega)}$  and 
$\frac{\ln(\Omega/\mu)}{\ln(\Omega_0/\Omega)}$
have invariant limit distributions as the fixed-point is approached ($\Omega\to 0$). This, together with $\ln(\Omega_0/\Omega)\sim \sqrt{\xi}$, suggests  
that the access probability $p(\xi)$ from a cluster to another one in a distance $\xi$ scales as $p(\xi)\sim\lambda\sim e^{-a\sqrt{\xi}}$, and the death rate of the cluster with the minimal $\mu$ within a distance $\xi$ scales as $\mu(\xi)\sim be^{-c\sqrt{\xi}}$. Here, $a,b$ and $c$ are random constants depending on the particular realization of the environment and whose distributions are independent of $\xi$. 
We have thus 
\be 
\mathcal{P}(t)\sim \sum_ne^{-a_n\sqrt{\xi_n}}e^{-b_ne^{-c_n\sqrt{\xi_n}}t}.
\ee
As the first factor decreases while the second one increases with $\xi_n$, the dominant contribution will be determined by a trade-off between them. Regarding the constants $a_n,b_n$ and $c_n$ independent of $n$ and approximating the sum by an integral, one obtains that its saddle-point is at 
$e^{c\sqrt{\xi^*}}=\frac{cb}{a}t$, resulting in $\mathcal{P}(t)\sim (\frac{cb}{a}t)^{-a/c}$. 
Keeping in mind that the constants were, in fact, random variables, we expect the probability density of $\ln[1/\mathcal{P}(t)]$ to possess the scaling property
\be 
f(\ln[1/\mathcal{P}(t)],t)=(\ln t)^{-1}\tilde f\left(\frac{\ln[1/\mathcal{P}(t)]}{\ln t}\right)
\label{phenscale}
\ee 
for long times with some unknown scaling function $\tilde f(\delta)$.
This kind of scaling form can be interpreted in the way that $\mathcal{P}(t)$ 
decays for long times as an inverse power of the time but the power $\delta$ that characterizes a given environment at a given time is a random variable having 
a time-independent probability density $\tilde f(\delta)$ in the limit $t\to\infty$.  

Note, however, that the above theory does not describe the atypical environments in that the lifetime of the primary cluster is greater than $t$ and therefore $\mathcal{P}(t)=O(1)$. 
Nevertheless, the effective decay exponents $\delta(t)=\ln[1/\mathcal{P}(t)]/\ln t$ of these samples lie in the range $0<\delta(t)<O(1/\ln t)$, so the lower bound of the domain where the scaling is valid tends to zero as $t\to\infty$. 

It follows from Eq. (\ref{phenscale}) that the typical survival probability defined as 
\be 
\mathcal{P}_{\rm typ}(t)\equiv \exp{\overline{\ln \mathcal{P}(t)}}
\label{Ptyp}
\ee
decreases algebraically with the time 
\be 
\mathcal{P}_{\rm typ}(t)\sim t^{-\delta_{\rm typ}},
\label{deltatyp}
\ee
where the decay exponent is given by the first moment of the distribution of $\delta$: 
\be 
\delta_{\rm typ}=\int_0^{\infty}\delta\tilde f(\delta)d\delta.
\ee 

Below the critical point, in the Griffiths phase, the correct scaling variables that have invariant limit distributions are $\Omega^{1/z}\ln(\Omega/\lambda)$ and $\mu/\Omega$. 
This, together with the relation $\Omega\sim \xi^z$ leads to the following form of the survival probability:
\be 
\mathcal{P}(t)\sim \sum_ne^{-a_n\xi_n}e^{-b_n\xi_n^{-z}t}.
\ee
Assuming again that the constants are independent of $n$, the saddle-point of the corresponding integral is at $\xi^*=\left(\frac{bzt}{a}\right)^{1/(1+z)}$, which results in
$\mathcal{P}(t)\sim\exp\left[-a(1+1/z)\left(\frac{bzt}{a}\right)^{1/(1+z)}\right]$. 
According to this, one expects the asymptotic scaling form for the probability density of $\ln[1/\mathcal{P}(t)]$
\be 
f_G(\ln[1/\mathcal{P}(t)],t)=t^{-\frac{1}{1+z}}\tilde f_G\left(\frac{\ln[1/\mathcal{P}(t)]}{t^{1/(1+z)}}\right) 
\label{gpscale}
\ee 
in the Griffiths phase.
The rescaled survival probabilities of atypical environments with $\mathcal{P}(t)=O(1)$, which are out of the validity of the above scaling arguments, are  $O(t^{-1/(1+z)})$ and, consequently, lie again in a domain that shrinks to zero as $t\to\infty$. 
The form of $\mathcal{P}(t)$ written in Eq. (\ref{ptsum}), as will be discussed in details in section \ref{auto}, is similar to that of the (imaginary-time) spin-spin autocorrelation function of the random transverse-field Ising chain. 
In the latter problem, based on a phenomenological picture of the Griffiths-McCoy phase, the scaling function has been obtained by Young in the form 
\be 
\tilde f_G(x)=c(cx)^{1/z}\exp\left[-\frac{z}{z+1}(cx)^{(1+z)/z}\right],
\label{fG}
\ee
where $c$ is a non-universal scale factor \cite{young} . 
According to Eq. (\ref{gpscale}), the typical survival probability follows a 
stretched exponential decay 
\be 
\mathcal{P}_{\rm typ}(t)\sim e^{-Ct^{1/(1+z)}}, 
\ee
where $C=\int_0^{\infty}x\tilde f_G(x)dx$. 

\section{The end-point survival probability}
\label{endpoint} 

\subsection{Calculation by the SDRG}
\label{surv1}

We start our analysis with the case of a semi-infinite chain, where, initially, an individual is placed on the first lattice site. The calculation of the corresponding survival probability $\mathcal{P}_1(t)$ in the frame of the SDRG method is simpler than that on an infinite lattice and so that its scaling function can obtained analytically.  

As a first step, we construct a coarse-grained model that consists of the clusters appearing in the representation of $\mathcal{P}_1(t)$ given in Eq. (\ref{ptsum}) by the SDRG procedure as follows. 
During the procedure, the first lattice site may become part of a cluster, which is the leftmost one, and, at some rate $\Omega=\Omega_1=\tilde\mu_1$, this cluster (called the primary one) will be decimated out. 
The next cluster appearing in the representation in Eq. (\ref{ptsum}) will be 
the active cluster that is next to the primary one at the scale $\Omega=\Omega_1$ and, after decimating out the primary cluster this will be the leftmost one. 
The effective reproduction rates between the primary and the secondary cluster are the rates $\tilde\nu_1,\tilde\kappa_1$ recorded at $\Omega=\Omega_1$. 
Then, at some lower rate $\Omega=\Omega_2<\Omega_1$, the leftmost cluster (the secondary one) will also be decimated. The rate $\tilde\mu_2=\Omega_2$ and the reproduction rates $\tilde\nu_2,\tilde\kappa_2$ toward the active cluster next to it are again recorded.  
Repeating this procedure ad infinitum, we obtain a coarse-grained contact process with a sequence of rates $\{\tilde\mu_n,\tilde\nu_n,\tilde\kappa_n\}_{n=1}^{\infty}$, which consists of much less sites than the original system. 
This can be quantified by calculating the mean number $n$ of sites of the coarse-grained model generated as the logarithmic rate $\Gamma=\ln(\Omega_0/\Omega)$
is increased from $\Gamma_i\gg 1$ to $\Gamma_f>\Gamma_i$. 
When $\Gamma$ changes by an infinitesimal amount $d\Gamma$ during the renormalization, the change in $n$ is given by the probability that the actually leftmost cluster is decimated out, which is $P_{\Gamma}(0)d\Gamma$, where 
$P_{\Gamma}(\beta)$ denotes the probability density of logarithmic death rate at the scale $\Gamma$. The decimation leaves the distribution of the death rate on the leftmost site identical to that of a bulk site \cite{fisher}, therefore we can use the fixed point solution in Eq. (\ref{fpsol}), yielding
\beqn
n=
\int_{\Gamma_i}^{\Gamma_f}P_{\Gamma}(0)d\Gamma=
\int_{\Gamma_i}^{\Gamma_f}\Gamma^{-1}d\Gamma=\ln\frac{\Gamma_f}{\Gamma_i}
\eeqn 
in the critical point. This is a double-exponential dependence of $1/\Omega_f$ on $n$, showing that the death rates $\Omega_n=\tilde\nu_n$ of subsequent sites of the coarse-grained model are well separated and $\tilde\mu_{n+1}/\tilde\mu_n\to 0$ as $n\to\infty$. (We mention that, according to Eq. (\ref{sp}), the probability distribution of $\Gamma\equiv\ln\frac{\Omega_0}{\tilde\mu_{n+1}}$ given that the previous death rate is $\tilde\mu_n$ is $P_>(\Gamma)=\Gamma_i/\Gamma$, where $\Gamma_i\equiv\ln\frac{\Omega_0}{\tilde\mu_{n}}$.)   
Using $\Gamma\sim\sqrt{\xi}$, we can see that a segment from site $1$ to site $L\gg 1$ of the original lattice is reduced by the SDRG to an effective model consisting of $n\simeq \frac{1}{2}\ln L + {\rm const}$ sites on average, which is a vanishing fraction of $L$. 

In the Griffiths phase, we obtain by similar calculations 
\beqn
n=
\int_{\Gamma_i}^{\Gamma_f}P_{\Gamma}(0)d\Gamma=
\frac{1}{z}(\Gamma_f-\Gamma_i)=\frac{1}{z}\ln\frac{\Omega_i}{\Omega_f},
\eeqn
where Eq. (\ref{pnullgp}) was used. Here, $\Omega_f$ decreases exponentially with $n$ and, according to Eq. (\ref{sps}), the ratio $r\equiv\tilde\mu_{n+1}/\tilde\mu_n$ has the limit distribution 
\be 
P_<(r)=r^{1/z}
\label{rdist}
\ee
as $n\to\infty$.    
Using $\Omega\sim \xi^{-1/z}$, we obtain here $n\simeq \ln L + {\rm const}$ for large $L$.

If one is interested in the long-time behavior of the model, a further simplification can be done. 
Consider two adjacent sites $n$ and $n+1$ of the coarse-grained model and 
assume that site $n+1$ is activated. If it is deactivated, which occurs with a rate $\tilde\mu_{n+1}$, it can be reactivated through site $n$ with a probability $\frac{\tilde\kappa_n}{\tilde\mu_{n+1}+\tilde\kappa_n}\frac{\tilde\nu_n}{\tilde\mu_{n}+\tilde\nu_n}<\frac{\tilde\nu_n}{\tilde\mu_{n}}\ll 1$, which is typically vanishing with increasing $n$.  
Therefore, once the activity has reached site $n$, the later history of the process is hardly influenced by sites with indices smaller than $n$ (which have much larger death rates) and, for large $n$, the process reduces to a random walk model. 
In this simplified model, the walker, the position of which represents the rightmost site that have been activated, either makes a step from site $n$ to site $n+1$ with a rate $\tilde\nu_n$, or dies (i.e. goes to an absorbing site) with a rate $\tilde\mu_n$. The survival probability in the original problem is then given by the probability that the walker starting on site $1$ is alive up to time $t$.  
As it is shown in Appendix \ref{mortal}, this can be explicitly given in terms of the jump rates
$\{\tilde\mu_{n},\tilde\nu_{n}\}_{n=1}^{\infty}$ of the walk as
\be
\mathcal{P}_1(t)=\sum_{n=1}^{\infty}a_ns_ne^{-\omega_nt},
\label{prw}
\ee 
where 
\beqn 
\omega_n &=&\tilde\mu_{n}+\tilde\nu_{n}, \\
a_n&=&\prod_{i=1}^{n-1}\frac{\tilde\nu_i}{\omega_i-\omega_n}, \\
s_n&=&1+\sum_{i=n}^{\infty}\prod_{j=n}^{i}\frac{\tilde\nu_j}{\omega_{j+1}-\omega_n}.
\label{sn}
\eeqn
This is of the form given in Eq. (\ref{ptsum}), and now we have an algorithm at hand for calculating the parameters appearing there from those of the original model in the frame of the SDRG method.  
The following heuristic way of obtaining a simple approximation to the  expression in Eq. (\ref{prw}) will turn out to be useful. 
The probability of reaching site $n>1$ starting from site $1$ is 
\be
p_n=\prod_{i=1}^{n-1}\frac{\tilde\nu_i}{\omega_i}
\label{pn}
\ee
and, once the walker is on site $n$, it will stay there for at least time $t$ with the probability $e^{-\omega_nt}$. 
Combining these leads to a simple expression 
\be
\mathcal{P}_1^0(t)=\sum_{n=1}^{\infty}p_ne^{-\omega_nt},
\label{prw0}
\ee
which is different from Eq. (\ref{prw}) but, for long times, they are asymptotically equal, at least in the critical point. To see this, note that $\omega_n/\omega_i\to 0$ for $i<n$,  in the limit $n\to\infty$, therefore $a_n/p_n\to 1$ and $s_n\to 1$ in that limit; consequently,  $\mathcal{P}_1(t)/\mathcal{P}_1^0(t)\to 1$ if $t\to\infty$.    
In the Griffiths phase, $s_n\to 1$ still holds, while the ratio $p_n/a_n$ can be written as  
\be 
\frac{p_n}{a_n}=(1-r_{n-1}')(1-r_{n-2}'r_{n-1}')\cdots (1-r_1'r_2'\cdots r_{n-1}'),
\ee
where $r_n'\equiv\frac{\omega_{n+1}}{\omega_n}$. For large $n$,  $r_n'\to r_n$, 
and $r_n<1$ follows the distribution given in Eq. (\ref{rdist}). The ratio 
$p_n/a_n$ is therefore an $O(1)$ random variable with an $n$-independent distribution in the limit $n\to\infty$, and finally we obtain that  $\mathcal{P}_1^0(t)/\mathcal{P}_1(t)=O(1)$ for large $t$. But, an $O(1)$ factor in 
$\mathcal{P}_1(t)$ does not affect the scaling function $\tilde f(\delta)$  of $\delta\equiv\ln[1/\mathcal{P}_1(t)]/\ln t$, so, for determining $\tilde f(\delta)$, we can start out from the simpler $\mathcal{P}_1^0(t)$. 

The scaling function $\tilde f(\delta)$ in the critical point can be calculated as follows. 
For the sake of simplicity, we will not distinguish between $\omega_n=\tilde\mu_n+\tilde\nu_n$ and $\Omega_n\equiv\tilde\mu_n$ since they are asymptotically equal. 
Let us consider, in a given realization of the environment and for a fixed time $t$, the term $n=M$ in the expansion of $\mathcal{P}_1^0(t)$ for which 
\be 
\Omega_M<\Omega\equiv t^{-1}<\Omega_{M-1}. 
\ee
The index $M=M(t)$ defined in this way is a non-decreasing function of $t$.  
As it is pointed out in Appendix \ref{corrections}, the term 
\be 
T_M=p_Me^{-\Omega_M/\Omega}
\label{TM}
\ee
dominates the sum in Eq. (\ref{prw0}) in the sense that terms with $n>M$ are vanishing with increasing $M$ (or, equivalently, with increasing time $t$) compared to $T_M$. Furthermore, the contribution of terms with $n<M$ to $\ln[1/\mathcal{P}_1(t)]/\ln t$ is asymptotically vanishing compared to that of the dominant one outside the domain
\be 
\Gamma_{M-1}<\Gamma <\Gamma_{M-1} + O[\ln(\eta_{M-1}\Gamma_{M-1})]. 
\label{band}
\ee
Since $\Gamma$ falls in the domain $(\Gamma_{M-1},\Gamma_{M})$, the width of which is typically $O(\Gamma_{M-1})$, the probability that those terms give a non-negligible contribution is vanishing with increasing $M$. 

The scaling function $\tilde f(x)$ we are looking for is the limit distribution of the variable $-\ln T_{M(t)}/\Gamma(t)$ or, equivalently, that of $-\ln p_{M(t)}/\Gamma(t)$ since $e^{-\Omega_M/\Omega}=O(1)$. 
Using Eq. (\ref{pn}) and that, according to Eq. (\ref{fpsol}), the variable 
$\eta_i=-\ln(\frac{\tilde\nu_i}{\omega_i})/\Gamma_i$ has an (exponential) limit distribution in the fixed point, the desired distribution is that of    
\be 
x=\Gamma^{-1}(t)[\eta_1\Gamma_1+\eta_2\Gamma_2+\dots +\eta_{M(t)-1}\Gamma_{M(t)-1}]
\label{gammasum}
\ee 
in the limit $t\to\infty$. The direct calculation of $\tilde f(x)$ seems to be complicated as the different variables $\Gamma_m$ are not independent and, in addition to this, the number of terms for a given $t$ is random.  
But, keeping in mind that $p_{M(t)}$ is nothing but the access probability 
$p_{\Omega}$ to the actually leftmost active cluster at scale $\Omega=t^{-1}$ of the renormalization, $\tilde f(x)$ can be calculated easily by following the evolution of its distribution during the procedure.

When the renormalization starts, the variable $p_{\Omega}$ is set to an initial value of $O(1)$ . It will change only if the actually leftmost cluster having rates $\tilde\mu,\tilde\nu$ is decimated. If this occurs, $p_{\Omega}$  
transforms as 
\be 
\tilde p_{\Omega} =p_{\Omega}\frac{\tilde\nu}{\tilde\mu+\tilde\nu}\simeq p_{\Omega}\frac{\tilde\nu}{\tilde\mu}
\label{ptrans1}
\ee
as can be seen from Eq. (\ref{pn}).
Or, using logarithmic variables $K\equiv\ln(1/p_{\Omega})$, 
$\beta=\ln(\Omega/\mu)$ and $\zeta=\ln(\Omega/\nu)$, 
we have asymptotically 
\be
\tilde K\simeq K+\zeta. 
\label{Kbeta}
\ee
The corresponding probability densities of the logarithmic variables at the scale $\Gamma$ will be denoted by 
$B_{\Gamma}(K)$, $P_{\Gamma}(\beta)$, and  $R_{\Gamma}(\zeta)$. 
When $\Gamma$ is increased by the renormalization by an infinitesimal amount to $\Gamma+d\Gamma$, the probability that the leftmost site is decimated is $P_{\Gamma}(0)d\Gamma$ and the distribution of $K$ changes as 
\beqn 
B_{\Gamma+d\Gamma}(K)=B_{\Gamma}(K) - P_{\Gamma}(0)d\Gamma\times 
\nonumber \\
\times [B_{\Gamma}(K) -\int dK'B_{\Gamma}(K')R_{\Gamma}(K-K')],
\eeqn
where the convolution describes the transformation of $K$, according to the rule in Eq. (\ref{Kbeta}). 
This leads to the differential-equation 
\be 
\frac{\partial B_{\Gamma}}{\partial\Gamma}=
- P_{\Gamma}(0)[B_{\Gamma}(K) -\int dK'B_{\Gamma}(K')R_{\Gamma}(K-K')].
\label{diffeq}
\ee
Looking for a scaling solution of the form 
\be 
B_{\Gamma}(K)=r_0(\Gamma)\tilde f[Kr_0(\Gamma)],
\ee
and, using the fixed point solution given by Eqs. (\ref{fpsol}) and 
(\ref{pnullcp}), we obtain that $\tilde f(x)$ satisfies the equation
\be 
x\frac{d\tilde f(x)}{dx}=-e^{-x}\int_0^xdx'\tilde f(x')e^{x'},
\ee
which has a simple solution 
\be 
\tilde f(x)=e^{-x}
\label{exp}
\ee
in terms of the scaling variable $x=Kr_0(\Gamma)=K/\Gamma$. 
This function gives the limit distribution of the effective decay exponents $\delta$ of the survival probability in the critical point when the process was started near the end of the lattice. 
We have found that, as far as the critical behavior of the model is controlled by the IRFP, not only the decay exponent of the average given in Eq. (\ref{pavcp}) but also the scaling function $\tilde f(\delta)$ is universal, i.e. independent of the distribution of parameters of the model.   
The exponent characterizing the algebraic decay of the typical survival probability defined in Eq. (\ref{Ptyp}) is 
\be 
\delta_{\rm typ,1}=1.
\ee
The scaling contribution to the average survival probability is 
\be 
\overline{\mathcal{P}_{\rm 1, sc}(t)}
=\int_{O(1/\ln t)}^{\infty}t^{-\delta}\tilde f(\delta)d\delta\sim (\ln t)^{-1},
\ee
which is of the same order of magnitude as the contribution coming from atypical samples, see Eq. (\ref{pavcps}). 

In the Griffiths phase, the scaling function of 
$Kr_0(\Omega)=KC\Omega^{1/z}$, where $C$ is a non-universal constant is still given by Eq. (\ref{exp}). However, it turns out that, here, the terms with $n<M$ can no longer be neglected compared to $T_M$, leading to that the scaling of the survival probability is characterized by a different power ($\frac{1}{1+z}$ rather than $\frac{1}{z}$) of $\Omega=t^{-1}$, see the phenomenological result in Eq. (\ref{gpscale}).

\subsection{Numerical results}
\label{num1}

\begin{figure}
\includegraphics[width=8cm]{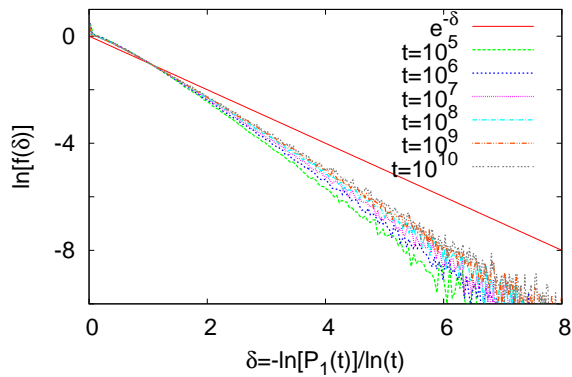}
\caption{\label{dcps} (Color online) Histogram of the rescaled survival probability in a semi-infinite system for different times in the estimated critical point. Data were obtained by the numerical SDRG procedure. The solid line is the limit distribution obtained analytically.}
\end{figure}

We have implemented the SDRG procedure numerically in systems of size $L=10^4$ and calculated the survival probability given by Eq. (\ref{prw}) up to $t=10^{10}$ in $10^5$ different random environments. At this time scale, $\mathcal{P}_1(t)$ is practically not affected by the other end of the lattice. 
The transition rates were 
\be
\mu_i=\frac{1}{1+r_i}, \quad \nu_i=\kappa_{i-1}=\frac{1}{2}\frac{r_i}{1+r_i}, 
\label{rates}
\ee
where the random variables $r_i$ were drawn from a uniform distribution in the interval $(0,l)$, the upper boundary $l$ being the control parameter of the phase transition. The critical point was estimated by the condition that the difference between the typical values of the last remaining reproduction and death rates obtained by the SDRG goes to a constant, i.e. 
$\overline{\ln\tilde\lambda(L)}-\overline{\ln\tilde\mu(L)}\to {\rm const}$ as $L\to\infty$. 
The distributions of the rescaled survival probability in the estimated critical point $l_c=9.3394(1)$ are shown in Fig. \ref{dcps}. 
The distributions for finite $t$ tend extremely slowly to the limit distribution, which can be understood since, according to Eq. (\ref{band}), the subdominant terms with indices $n<M$ are non-negligible with a probability of 
$O(\frac{\ln\ln t}{\ln t})$ which tends to zero very slowly. The width of this zone, $O[\ln(\eta_{M-1}\Gamma_{M-1})]$, is larger for samples with larger $\eta_{M-1}$ and, according to Eq. (\ref{gammasum}), in these samples, $\delta$ is relatively large, a well. Therefore, deviation from the limit distribution is more significant in the large-$\delta$ region.  

We have also considered a simplified version of the renormalization procedure, which leads to the same universal properties at the critical point as the original one, but it has the advantage that the location of the critical point is known exactly. Dropping, namely, the factor $2$ in the transformation rule of $\lambda$ in Eq. (\ref{lambda_appr}), it will be identical to that of $\mu$ in Eq. (\ref{mu_appr}). 
Applying these symmetrized transformation rules for the variables $\mu_i$ and $\lambda_i$ from the beginning of the decimation procedure, an initially identical distribution of these rates will remain identical, but this is an exclusive property of the critical fixed point, see Eq. (\ref{fpsol}). We have calculated the survival probability by the symmetrized procedure, using uniform distribution of the rates $\lambda$ and $\mu$ and we have simply substituted $\lambda$ for $\nu$ in the expression of the survival probability in Eq. (\ref{prw}). 
\begin{figure}
\includegraphics[width=8cm]{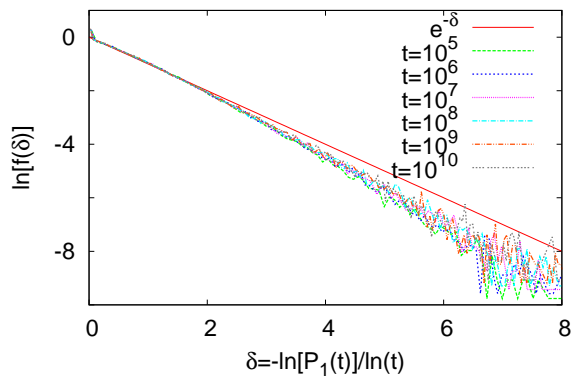}
\caption{\label{dis} (Color online) Histogram of the rescaled survival probability in a semi-infinite system for different times obtained by the symmetrized renormalization rules described in the text.}
\end{figure}
Although the limit distribution is the same, the finite-time distributions depend on the initial rates and they are also influenced by the differences in the renormalization rules, as can be seen by comparing Fig. \ref{dcps} and \ref{dis}.
The data obtained by the symmetrized method are closer to the limit distribution, which may also be attributed to that these data are free from the error of locating the critical point. 

\begin{figure}
\includegraphics[width=8cm]{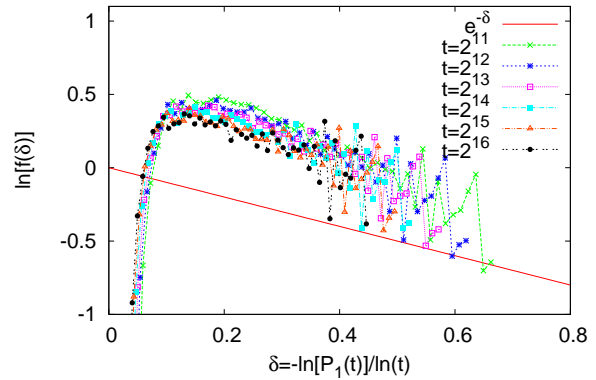}
\caption{\label{ptds} (Color online)  Histogram of the rescaled survival probability in a semi-infinite system for different times obtained by Monte Carlo simulations in the estimated critical point.}
\end{figure}

We have also performed Monte Carlo simulations of the discrete time variant of the process with rates given in Eq. (\ref{rates}). The survival probability was measured in $10^5$ different samples by performing $2000$ runs per sample. 
The critical point has been located by finding the point where the relationship between the average survival probability and the average number of individuals is algebraic \cite{vojta}, which gives the estimate $l_c=8.80(3)$. Note that the location of the critical point estimated by the SDRG method is different from this value, owing to the approximative nature of that method at finite scales. The number of runs per sample limits the resolution of the measurement of $\mathcal{P}_1(t)$, therefore the range of $\delta$ is much narrower here than for the SDRG method. The times available by the simulation lag behind those of the SDRG, as well, and, although the distributions are rather far from the limiting one, they seem to approach it for increasing time, as can be seen in Fig. \ref{ptds}.      

\section{The bulk survival probability}
\label{bulk}

Next, we turn to the problem of calculating the survival probability in an infinite system. This question is far more complicated than the 
calculation of $\mathcal{P}_1(t)$, and we could not obtain the complete form of the corresponding scaling function. Nevertheless, it can still be determined numerically by the SDRG procedure, and its limiting behaviors can be guessed. 

\begin{figure}
\includegraphics[width=7cm]{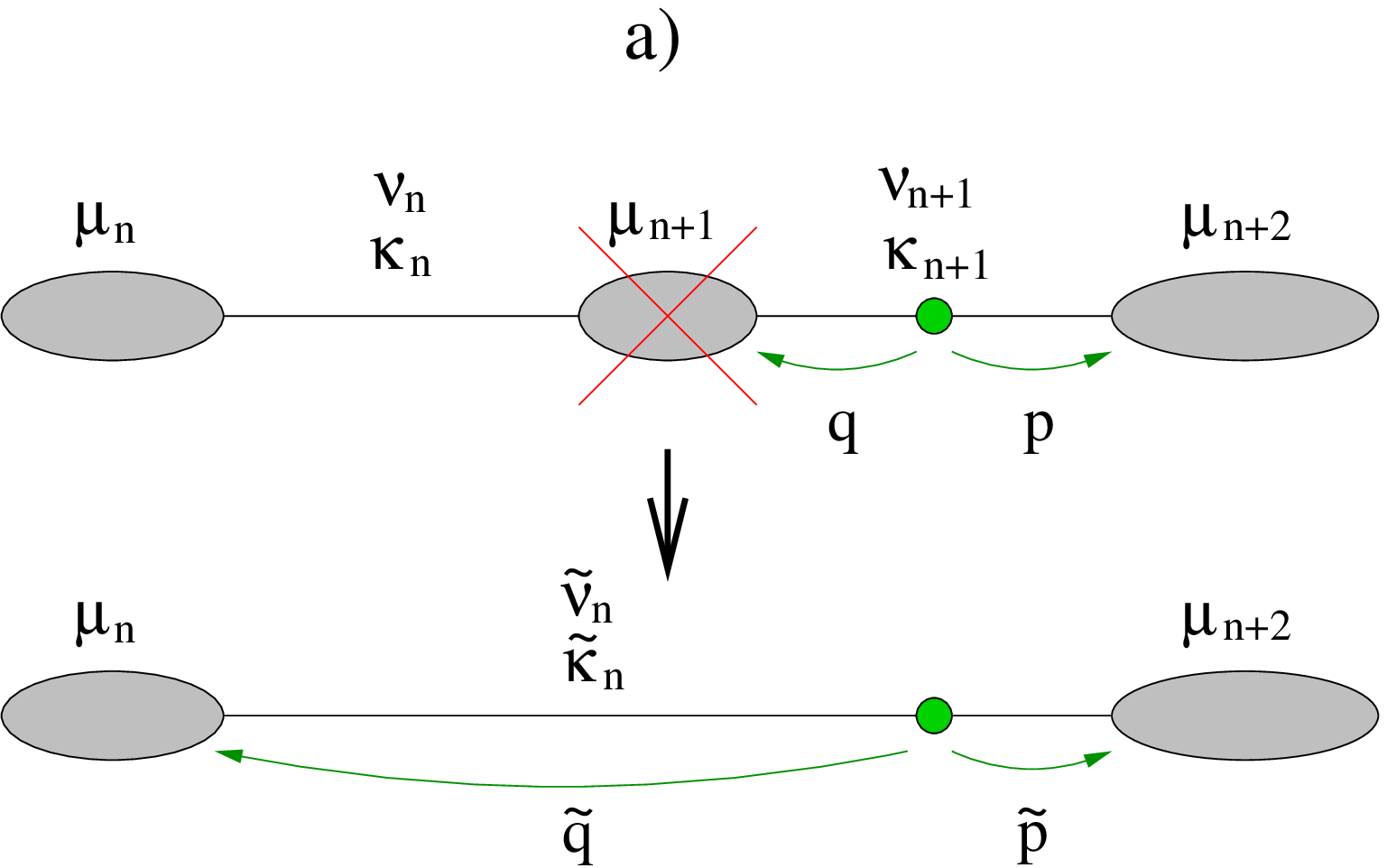}
\includegraphics[width=7cm]{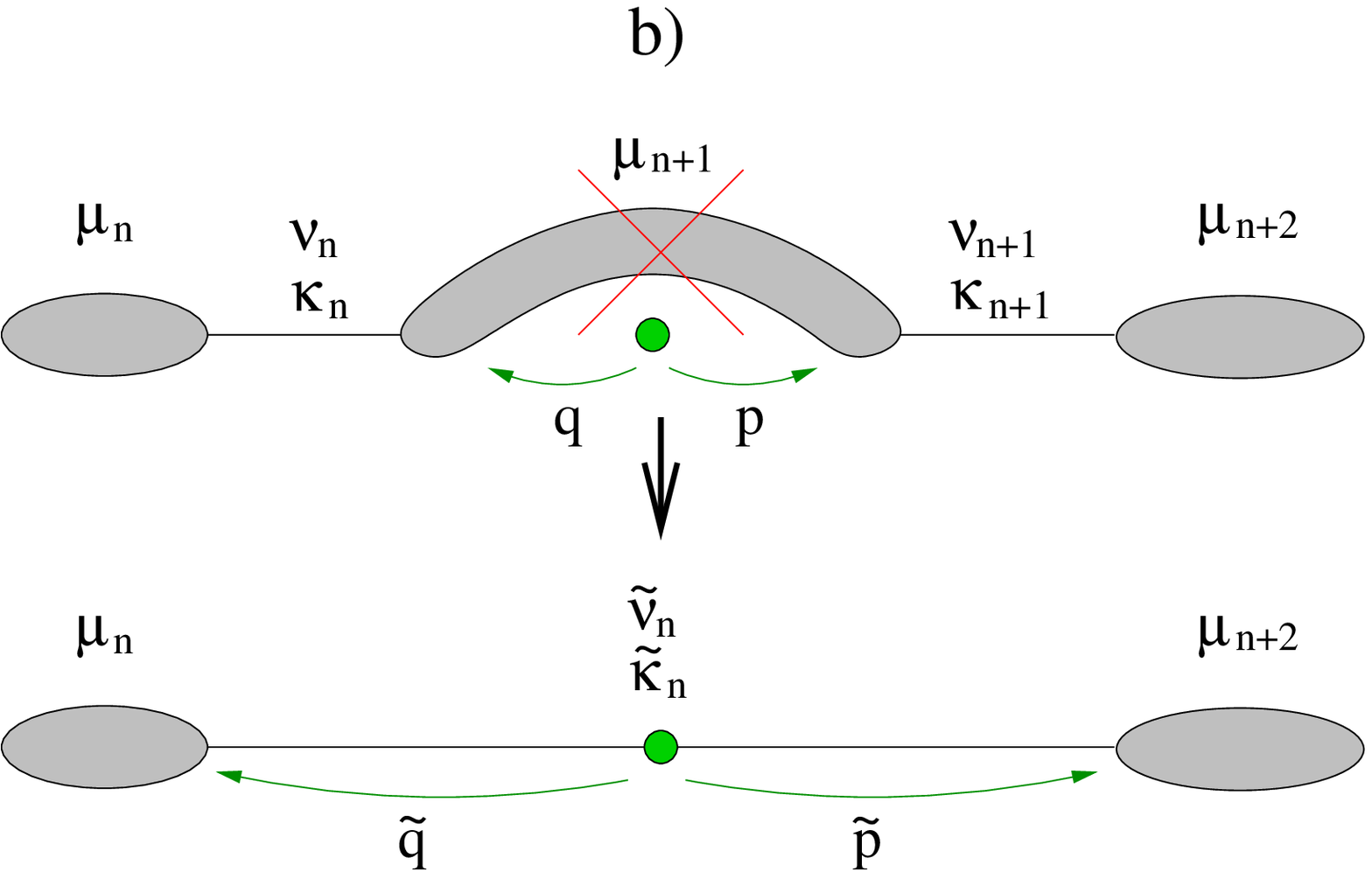}
\caption{\label{bulk1} (Color online) Illustration of the renormalization of access probabilities $p$ and $q$ when the target clusters are distinct (a) and when the starting site is surrounded by one single target cluster at a given stadium of the decimation procedure (b).}
\end{figure}

\begin{figure}
\includegraphics[width=8cm]{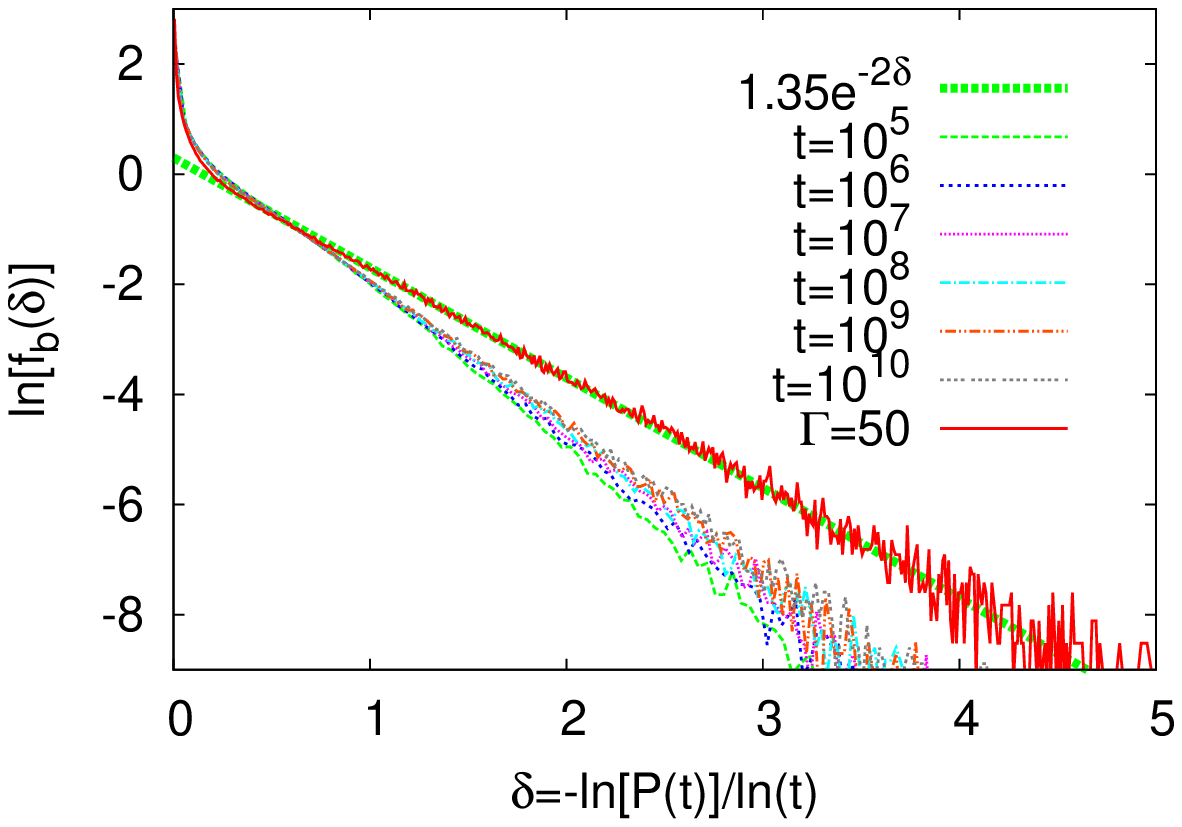}
\includegraphics[width=8cm]{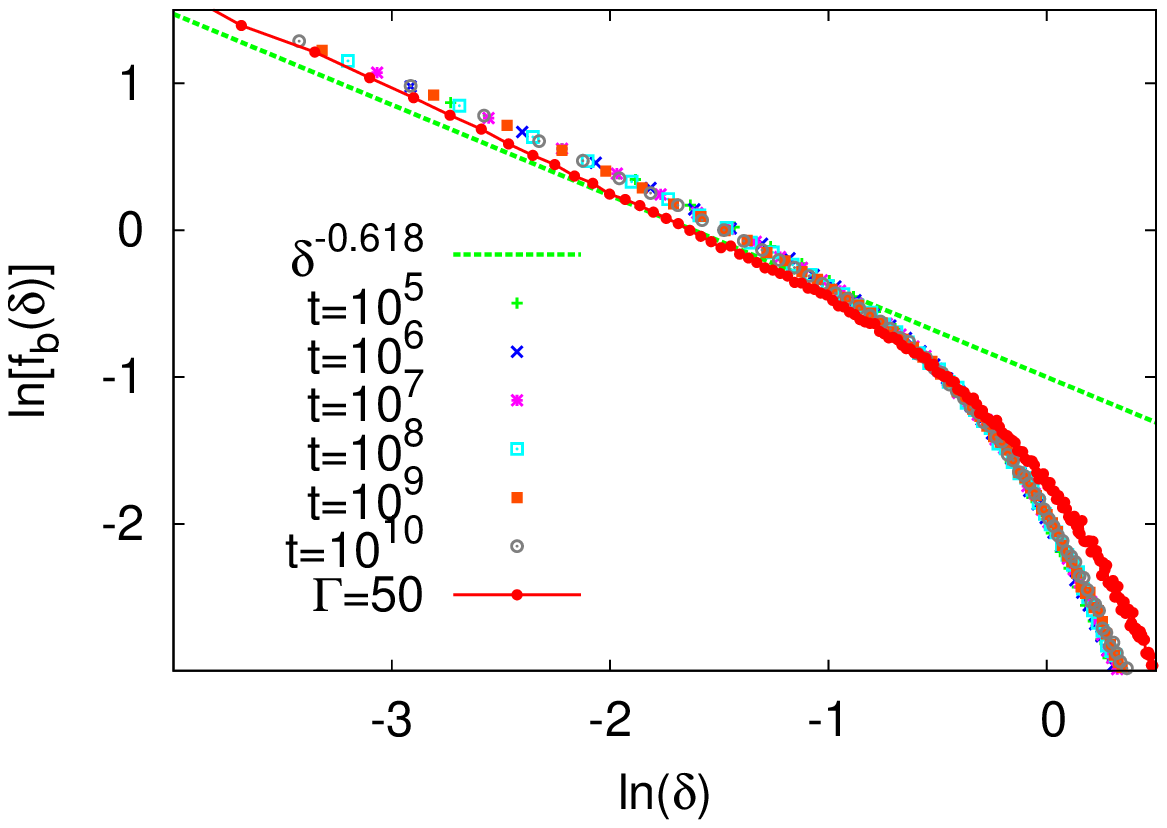}
\caption{\label{bulk2} (Color online) 
Top. Histogram of the rescaled survival probability for different times in the estimated critical point. Data were obtained by the numerical SDRG procedure.
The histogram of the variable $-\ln(p_{\Omega}+q_{\Omega})/\Gamma$ calculated at $\Gamma=50$ is also shown. The straight line corresponds to an exponential function $\sim e^{-2\delta}$. 
Bottom. The same data as above plotted against $\ln(\delta)$. Here,  
the solid line corresponds to $\sim\delta^{-0.618}$. 
}
\end{figure}

The infinite system can be coarse-grained by the SDRG procedure, then can be further reduced to a mortal random walk problem in the way similar to the semi-infinite one described in the previous section.
Here, however, from a given cluster, the activity can get, in general, to two further clusters (one on its right and one on its left) rather than one and, in some cases, these two may be merged later during the decimation procedure, so the corresponding mortal random walk problem is defined on a random network rather than on a one-dimensional lattice.  
In this more complicated case, we could not find an exact expression of the survival probability of the mortal walker, but the asymptotically correct expression given in Eq. (\ref{prw0}) can still be applied.  
In order to evaluate this expression, one follows up the access probability to the closest active cluster on the r.h.s. and on the l.h.s. of the starting site denoted by $p$ and $q$, respectively, during the decimation procedure, and records them together with the decay rate $\omega\approx \tilde\mu$ whenever either of these clusters is decimated out. 
Concerning the transformation of $p$ and $q$, two different situations must be distinguished, as illustrated in Fig. \ref{bulk1}. 
One possibility is that the two ``target'' clusters labeled by $n+1$ and $n+2$ in the figure are distinct (at least at the scale $\Omega$ under consideration).  If the one with label $n+1$ is decimated out, 
its decay rate $\omega_{n+1}=\mu_{n+1}+\kappa_n+\nu_{n+1}$ is recorded and it 
is replaced by a new target cluster (the one with label $n$) having the  
access probability  
\be
\tilde q \simeq q\frac{\kappa_n}{\mu_{n+1}+\kappa_n}.
\ee
But, owing to the elimination of cluster $n+1$, the access probability to cluster $n+2$ is also modified, since besides the ``direct'' route to $n+2$ with probability $p$, another one via cluster $n+1$ opens, which is taken into account by a renormalized access probability  
\be 
\tilde p \simeq p + q\frac{\nu_{n+1}}{\mu_{n+1}+\nu_{n+1}}. 
\ee
The other possibility is that the two target clusters are merged during the renormalization and the starting site is then surrounded by one single active cluster, which can be reached via two different routes, as illustrated in Fig. \ref{bulk1}. If it is decimated out, there will again be two distinct target clusters with access probabilities 
\be
\tilde q \simeq (q+p)\frac{\kappa_n}{\mu_{n+1}+\kappa_n},  
\qquad 
\tilde p \simeq (q+p)\frac{\nu_{n+1}}{\mu_{n+1}+\nu_{n+1}}. 
\ee
This transformation scheme looks much more complicated than that of semi-infinite system in Eq. (\ref{ptrans1}). Furthermore, the variables $q$ and $p$ become correlated with the reproduction rates $\nu$ and $\kappa$ on the link connecting the two actual target clusters during the renormalization, as can be seen by inspecting one decimation step as shown in Fig. \ref{bulk1}a. 
For these reasons, we could not treat the problem of calculating the scaling function of the survival probability analytically. 
Instead, we have computed $\mathcal{P}(t)$ numerically using the rates given in Eq. (\ref{rates}) in the same way as described in the previous section, except that the starting site was far from the boundaries (in the middle of the system).We have also determined the distribution of the variable $x=-\ln(p_{\Omega}+q_{\Omega})/\Gamma$, where $p_{\Omega}$ and $q_{\Omega}$ are the access probabilities to the actual target cluster(s) at the scale $\Omega$. 
Following the arguments presented in the previous section, one concludes that 
the probability density of $x$ in the large-$\Gamma$ limit gives the scaling function $\tilde f_b(x)$ of the distribution of the survival probability, where the subscript 'b' stands for 'bulk'.  
The distribution of $x$ is found to converge rapidly to the limit distribution, 
and, the latter being universal, we have used the symmetrized renormalization rules for computing $x$. The data obtained at $\Gamma=50$, shown in Fig. \ref{bulk2}, are expected practically not being distinguishable from the limit distribution $\tilde f_b(x)$. 
The rescaled distributions of the survival probability at finite times, shown in the same figure, converge very slowly to the limit distribution for the reason discussed in the previous section.

Although the analytical form of $\tilde f_b(x)$ is not known, an upper bound on the corresponding distribution function $P_{>,b}(x)=\int_x^{\infty}\tilde f_b(x')dx'$ can be derived by simple arguments as follows. 
Assume that the process starts with two individuals on the neighboring sites $0$ and $1$ rather than with a single one, which is an unimportant modification regarding the asymptotic properties such as the scaling function 
$\tilde f_b(x)$. 
Now, if we block the transmission of activity between two semi-infinite parts
of the system by setting $\nu_0=\kappa_0=0$, 
the survival probability will obviously be reduced for any $t$.
The scaling function in the resulting system, which consists of two independent semi-infinite chains is easy to calculate. 
Denoting the access probabilities in the two parts by 
$p_{\Omega}$ and $q_{\Omega}$ at the scale $\Omega$, the scaling function 
$\tilde f_{\rm cut}(x)$ of this system will be the limit distribution of 
\be 
x=-\frac{\ln[p_{\Omega}+q_{\Omega}]}{\Gamma}. 
\ee
Since the distribution of access probabilities broadens without limits as $\Gamma$ increases, their sum will asymptotically be equal to the greater one of them, 
$p_{\Omega}+q_{\Omega}\simeq\max\{p_{\Omega},q_{\Omega}\}$, leading to 
\be 
x\simeq -\frac{\ln(\max\{p_{\Omega},q_{\Omega}\})}{\Gamma}=
\min\{x_p,x_q\},
\ee
where $x_p$ and $x_q$ denote the corresponding scaling variables in the two independent parts, each having the distribution $\tilde f(x)=e^{-x}$.  
We obtain therefore $\tilde f_{\rm cut}(x)=2e^{-2x}$ and the upper bound 
\be 
P_{>,b}(x)\equiv\int_x^{\infty}\tilde f_b(x')dx'<e^{-2x}.
\ee
Interestingly, the scaling function $\tilde f_b(\delta)$ fits well to the exponential function $Ce^{-2\delta}$ for not too small values of $\delta$, as can be seen in Fig. \ref{bulk2}, suggesting that the scaling variable $\delta$ in samples in which it is large (the survival probability is small) is composed of two almost independent contributions.   
For $\delta$ approaching zero, the scaling function seems to diverge, as can be seen in the lower part of Fig. \ref{bulk2}, as an inverse power of $\delta$.  
Regarding that, in the semi-infinite system, the scaling and non-scaling contributions to the average survival probability were found to be of the same order of magnitude, a similar outcome for the bulk case would require a divergence of 
the form   
\be 
\tilde f_b(\delta)\sim \delta^{-1+x_b/\psi},
\ee
as $\delta\to 0$. 
The numerical data are compatible with this form, in particular with the value of the exponent $1-x_b/\psi\approx 0.618$. 

The typical survival probability defined in Eq. (\ref{Ptyp}) decreases algebraically with the time, and the universal decay exponent $\delta_{\rm typ,b}$, which is the first moment of $\tilde f_b(\delta)$ is, according to our estimates,
\be 
\delta_{\rm typ,b}=0.34(1).
\ee

\begin{figure}
\includegraphics[width=8cm]{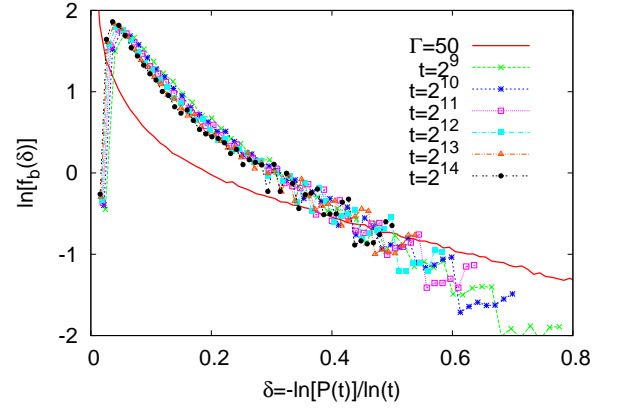}
\caption{\label{bulksim} (Color online) 
Histogram of the rescaled survival probability for different times obtained by Monte Carlo simulations in the critical point. The solid curve is the limit distribution obtained by the SDRG method. 
}
\end{figure}

We have also measured the distribution of the survival probability by Monte Carlo simulations in the same way as described in the previous section, except that, here, periodic boundary conditions were used. 
The data obtained in the estimated critical point, shown 
in Fig. \ref{bulksim}, are far from the universal limit distribution owing to strong finite-time corrections. 
One can, however, observe the signs of a developing singularity at $\delta=0$ even at the time-scales accessible by simulations.  

\begin{figure}
\includegraphics[width=8cm]{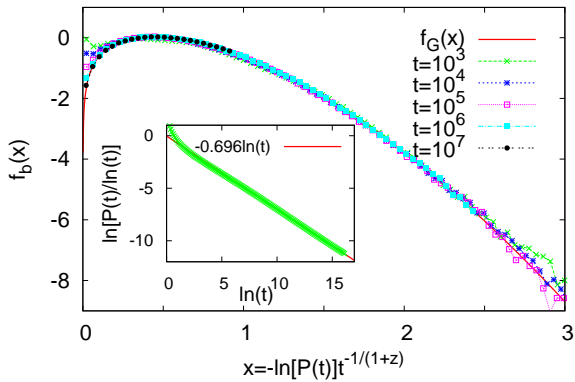}
\caption{\label{dgp} (Color online) 
 Histogram of the rescaled survival probability for different times in the Griffiths phase. Data were obtained by the numerical SDRG procedure using the distribution of rates given in Eq. (\ref{rates}) with $l=7$. 
The solid curve is the phenomenological scaling function calculated in Ref. \cite{young} and given in Eq. (\ref{fG}) with the scale factor $c=1.82$.  
In the inset, the dependence of the average survival probability on time is shown. The straight line is a linear fit to the data, the slope of which is, according to Eq. (\ref{pavgp}), $1/z$.  
}
\end{figure}

Finally, we have also determined the distribution of the survival probability in the Griffiths phase numerically by the SDRG method. 
Using the distribution of rates given in Eq. (\ref{rates}) with $l=7$, we have calculated the average survival probability and, using Eq. (\ref{pavgp}), estimated the dynamical exponent by a linear fit, as shown in the inset of Fig. \ref{dgp}, which gives $z=1.44(1)$.   
The distributions of the survival probability rescaled with this estimate fit well to the phenomenological scaling function in Eq. (\ref{fG}), as can be seen in the figure.  

\section{Survival probability of random walks in random environments}
\label{rwre}

A dynamical process simpler than the contact process is the random walk in random environment \cite{sk,bouchaud}, where the time-dependent survival probability, as we shall see, is closely related to that of the contact process on a semi-infinite lattice. 
Let us consider continuous-time random walks on the non-negative integers with
independent, identically distributed random jump rates $u_i$ and $v_{i-1}$ from $i$ to $i+1$ and $i-1$, respectively. 
Assume, furthermore, that $u_0=0$, meaning that site $0$ is an absorbing one, and the walker starts at $t=0$ from site $1$. 
We are interested in the probability $\mathcal{P}(t)$ that the walker does not visit the absorbing site up to time $t$ in a fixed random environment. 
The contribution from atypical random environments with $\mathcal{P}(t)=O(1)$
to the average is known to decrease as \cite{fdm,ir,cd}
\be 
\overline{\mathcal{P}_{\rm atyp}(t)}\sim (\ln t)^{-1}, 
\ee  
but, to our knowledge, the scaling function of the distribution of the survival probability has not been calculated in this model yet.

As opposed to the random walk problem considered in appendix \ref{mortal}, here, we could not find a closed expression of the survival probability in terms of the jump rates. 
But, applying the SDRG method to this problem, an approximate form, which is expected to yield the correct scaling function, can be constructed in the way it has been done to obtain Eq. (\ref{prw0}).
Instead of an SDRG treatment based on the energy landscape \cite{fdm}, we need here an alternative formulation in terms of the transition rates \cite{jsi,mg}, which is formally closer to the SDRG scheme of the contact process.
In this procedure, first, the largest rate $\Omega=\max\{u_i,v_i\}$ in the system is found. If it is $\Omega=u_n$, then the sojourn time $1/(u_n+v_{n-1})$ on site $n$ is small compared to other sites, therefore it is eliminated and site $n-1$ and $n+1$ become directly accessible with the effective jump rates 
\beqn
\tilde u=\frac{u_{n-1}u_n}{\omega_n}\simeq u_{n-1}, \nonumber \\
\tilde v=\frac{v_{n-1}v_n}{\omega_n}\simeq\frac{v_{n-1}v_n}{u_n},
\label{uv}
\eeqn 
where $\omega_n=u_n+v_{n-1}$, and one obtains similar expressions in the case $\Omega=v_n$ by symmetry. This decimation step is then performed iteratively.  
Approaching the fixed point of the transformation, which is an IRFP, 
$v_{n-1}/\omega_n$ tends to zero, and the  
approximations in Eqs (\ref{uv}) will be asymptotically exact.  
In order to make the similarity between the above decimation scheme and that of the contact process clearer, let us interpret the decimation of a rate $u_n$ that site $n$ joins the site on its right hand side and they form a cluster with a reduced jump rate $\tilde v$ to the left, while, if a rate $v_{n-1}$ is selected for decimation then site $n$ is simply deleted. 
Notice that this renormalization scheme is formally identical with the symmetrized renormalization of the contact process described in section \ref{num1}
with the correspondences 
\be 
u_n\leftrightarrow\lambda_n, \qquad 
v_{n-1}\leftrightarrow\mu_n. 
\ee
We will show that the form of the survival probability is identical in the two problems, as well.  
The key quantity for constructing the time-dependent survival probability is the probability $\pi_{n,m}$ that, starting from site $n$, the walker reaches site 
$m$ (before it is trapped on the absorbing site). 
For fixed $m$ and different $n$, these hitting probabilities obey the recursion relations 
\be 
\pi_{n,m}=\frac{v_{n-1}}{\omega_n}\pi_{n-1,m} + \frac{u_{n}}{\omega_n}\pi_{n+1,m}, 
\label{recpi}
\ee
with the boundary condition $\pi_{m,m}=1$. 
This can be used to obtain the access probability $p_m\equiv\pi_{1,m}$ explicitly in the form of a Kesten variable, see e.g. Ref \cite{ir}, but, instead of this, we will keep track of $p_m$ during the SDRG procedure. 
Using Eq. (\ref{recpi}), one can show that, if a site (which is different from $1$ and $m$) is decimated, the access probability $\pi_{1,m}$ remains exactly invariant provided the effective rates are calculated according to Eq. (\ref{uv}). 
We can construct an approximate form of the survival probability in the frame of the SDRG procedure as follows. 
At the beginning of the renormalization, the access probability of site $1$ is $p_1=1$ and, at some scale $\Omega$, one of the rates emanating from this site is selected for decimation. 
If it is $u_1$, then, assuming $v_0\ll u_1=\Omega$, the walker, when leaving this site, will jump typically to the site on its right rather than to the absorbing one. Formally, site $1$ joins its right neighbor, forming a cluster (the leftmost one). 
Identifying the access probability of a cluster with that of its rightmost site, the access probability of the leftmost cluster will be $\tilde p_1=\frac{u_1}{\omega_1}p_1=\frac{u_1}{\omega_1}$ after the above decimation step.
In general, if the jump rate $\tilde u_1$ of the leftmost cluster is selected for decimation $\tilde p_1$ will pick up a factor 
$\frac{\tilde u_1}{\tilde\omega_1}$, which tends $1$ as the fixed point is approached, leaving the access probability asymptotically unchanged.  
If, however, the rate $\tilde v_0$ of the leftmost cluster is the maximal one, 
then the walker will typically jump from this cluster on the absorbing site. 
In this case, the leftmost cluster is eliminated, and the term
$\tilde p_1e^{-\tilde\omega_1t}$ will give the leading term of $\mathcal{P}(t)$.
The access probability of the active cluster next to the eliminated one, right after the decimation is 
\be 
\tilde p_2=\frac{\tilde u_1}{\tilde\omega_1}\tilde p_1,
\ee   
and it will not be essentially modified when this cluster is possibly merged with its right neighbor. 
When this cluster is decimated out at some scale $\Omega=\tilde v_1$, a term $\tilde p_2e^{-\tilde\omega_2t}$ is added to $\mathcal{P}(t)$, and this step is repeated whenever the actually leftmost cluster is decimated during the renormalization. 

We have thus obtained that the approximate form of the survival probability, which is expected to yield the correct scaling function, is identical to that of the contact process on a semi-infinite lattice (calculated by the symmetrized SDRG method), so we can make use of the results of section \ref{surv1}. 
The critical point of the contact process corresponds to jump rate distributions of the random walk problem that satisfy the relation $\overline{\ln u}=\overline{\ln v}$ \cite{sk}. 
In this case, we obtain that the scaling function $\tilde f_{\rm rwre}(x)$ of 
$x=\ln[1/\mathcal{P}(t)]/\ln t$ is
\be 
\tilde f_{\rm rwre}(x)=e^{-x}. 
\ee
The case  $\overline{\ln u}<\overline{\ln v}$, when the walker is forced to move toward the absorbing site, corresponds to the Griffiths phase of the contact process. Here, $\mathcal{P}(t)$ has the scaling property given in Eq. (\ref{gpscale}) with the scaling function in Eq. (\ref{fG}), and the dynamical exponent $z$ appearing there is the positive root of the equation 
$\overline{(u/v)^{1/z}}=1$ \cite{sk,dp}.

\section{Disordered quantum spin chains}
\label{quantum}
 
The earliest examples of an IRFP were provided by zero temperature phase transitions of certain random quantum spin chains, and the method of SDRG originates in this field \cite{mdh,fisher,im}. 
These are, at the same time, the models in which the details of an infinite randomness critical behavior are the most elaborated. 
Here, we shall point out that the same scaling functions obtained in the context of the contact process arise in the description of apparently different quantities of these models, as well.  
To be concrete, we consider the random transverse-field Ising chain (RTIC) defined by the Hamiltonian
\be 
\mathcal{H}=-\sum_iJ_{i}\sigma^x_i\sigma^x_{i+1}-\sum_ih_i\sigma^z_i,
\label{hamilton}
\ee
where $\sigma_i^{x}$ and  $\sigma_i^{z}$ are Pauli operators on site $i$, and the bonds $J_i$ and external fields $h_i$ are independent, identically distributed random variables.
If $\Delta\equiv\overline{\ln h}-\overline{\ln J}<0$, the model is in its ferromagnetic phase characterized by a non-zero spontaneous magnetization, which vanishes continuously as the quantum critical point at $\Delta=0$ is approached, and for $\Delta>0$, the model is paramagnetic.

In the SDRG scheme of this model \cite{fisher}, the largest coupling (either a bond or a field), which determines the energy scale $\Omega$, is looked for. 
If it is a bond, $\Omega=J_1$, furthermore $J_1\gg h_1,h_2$, then, according to perturbative calculations, the spins on site $1$ and $2$ will flip coherently 
and experience a transverse field of effective magnitude
\be 
\tilde h\simeq\frac{h_1h_2}{J_1}.  
\ee
While, if $\Omega=h_2\gg J_1,J_2$, the spin on site $2$ is pinned in the direction of the transverse field, 
therefore it is decimated and spins on site $1$ and $3$ are connected by an effective bond 
\be 
\tilde J\simeq\frac{J_1J_2}{h_2}.  
\ee
This SDRG scheme is apparently equivalent to that of the contact process in its symmetrized form with the correspondences 
\be 
J_i\leftrightarrow\lambda_i, \qquad 
h_i\leftrightarrow\mu_i. 
\ee

\subsection{Spontaneous magnetization}

In the RTIC, distributions of various quantities have been studied by the free-fermion technique \cite{ky,ijr,young,yr} or by the SDRG method \cite{fisher,fy}, among others that of the end-point spontaneous magnetization $m_1\equiv\lim_{H\to 0}\langle\sigma^x_1(H)\rangle$ in a semi-infinite chain. Here, $H$ is the magnitude of a magnetic field applied in the $x$-direction on the first spin only and $\langle\cdot\rangle$ denotes the ground state expectation value.  
The way of calculation of $m_1$ by Fisher \cite{fisher} is analogous to that of the survival probability of the contact process. To see this, we will now briefly recapitulate the initial part of the calculation.  For a given (small) $H$,  the decimation procedure is stopped at the energy scale $\Gamma_H=\ln(\Omega_0/H)$. If the surface spin is still active in a cluster at this scale, the magnetization $m_1(H)$ will be $O(1)$.  
If it is no longer active, it still has a small but non-zero magnetization due to the correlation with the leftmost active spin in the closest cluster, $m_1^c(H)\sim \langle \sigma_1^x\sigma_L^x\rangle$.  
When the primary cluster is decimated at the scale $\Omega=\tilde h_1$,
a simple, perturbative calculation gives that this correlation
is $m_1^c(H)\sim \tilde J_1/\tilde h_1$, where $\tilde J_1$ is the bond to the leftmost active spin. Then each further decimation of the actually leftmost cluster brings a factor $\tilde J_i/\tilde h_i$ to $m_1^c(H)$. 
This procedure results in an expression of $m_1^c(H)$ similar to that of the dominant term of $\mathcal{P}_1(t)$ in the contact process.
In the ferromagnetic phase $\Delta<0$, the probability that a spin belongs to a cluster that is never decimated during the renormalization is finite.
In particular, for the surface spin, it is $O(1/z')$, where $z'$ denotes the dynamical exponent in this phase.
Therefore $m_1^c(H)$ will consist of a finite number of factors even in the limit $H\to 0$ ($\Gamma_H\to\infty$) if $\Delta<0$. 
Furthermore, the contribution of decimated clusters to the spontaneous magnetization $m_1$ is vanishing and given solely by $m_1^c(H)$ in that limit.  
Approaching the critical point, $\Delta\to 0-$, the fraction of ``atypical'' samples with $m_1=O(1)$, being $O(1/z')$, vanishes. In this limit, an exponential distribution of the rescaled spontaneous magnetization $\ln[1/m_1]\frac{1}{z'}$ of typical samples has been derived in Ref. \cite{fisher} by analysing a sum of type given in Eq. (\ref{gammasum}). 

Instead of this last step, a simpler alternative way of obtaining 
the above scaling form is provided by the method presented in section \ref{surv1}, as follows.  
Notice that the spontaneous magnetization calculated in the above way in the ferromagnetic phase is analogous to the stationary survival probability, $\mathcal{P}_1(\infty)\equiv\lim_{t\to\infty}\mathcal{P}_1(t)$, in the active phase of the contact process. 
Taking into account that the functions $r_0(\Omega)$ and $p_0(\Omega)$ change roles in the active phase compared to the inactive one, the scaling variable in terms of which the solution to Eq. (\ref{diffeq}) is an exponential, will be $x=Kr_0(\Omega)\simeq K\frac{1}{z'}$, resulting in an exponential distribution of $\ln[1/\mathcal{P}_1(\infty)]\frac{1}{z'}$.

Similar to the end-point magnetization, one can consider the bulk spontaneous magnetization $m_b$ in response to a vanishing magnetic field applied on a given spin of an infinite chain. As opposed to $m_1$, the distribution of this quantity is not known.   
But it is easy to extend the correspondence described above to the bulk and to see that the distribution of the rescaled magnetization $x=\ln[1/m_b]\frac{1}{z'}$ in the limit $\Delta\to 0-$ is given by the function $\tilde f_b(x)$, which has been numerically studied in section \ref{bulk}.

\subsection{Imaginary-time autocorrelations}
\label{auto}

The spin-spin autocorrelation
\be 
C_n(\tau)=\langle\sigma_n^x(0)\sigma_n^x(\tau)\rangle
\label{Cn}
\ee
as a function of the imaginary time $\tau=it$ in disordered quantum spin chains has been the subject of several works \cite{young,ky,ri,ijr} and shows a behavior similar to that of the survival probability in the contact process. 
According to a phenomenological scaling theory, in the critical point of the RTIC, its average decays with $\tau$ as 
\be 
\overline{C_n(\tau)}\sim (\ln\tau)^{-x_b/\psi}
\ee
in an infinite system, whereas, on the surface site $n=1$ of a semi-infinite chain, $\overline{C_1(\tau)}\sim (\ln\tau)^{-1}$ \cite{ri}. 
Furthermore, numerical analyses of the distribution of $C_n(\tau)$
based on the free-fermion technique show a data collapse of the 
scaling variable $\ln[1/C_n(\tau)]/\ln\tau$ in the critical point \cite{ky,ijr} 
and of the variable $\ln[1/C_n(\tau)]/\tau^{1/(1+z)}$ in the paramagnetic Griffiths phase \cite{young,ijr}. The latter behavior is predicted by a phenomenological theory, in the frame of which the scaling function has been calculated \cite{young}. 

Unlike in the Griffiths phase, the form of the scaling function unknown in the critical point. 
By applying the SDRG method we will show that it can be related to that of the survival probability of the contact process. 
Let us start by rewriting $C_n(\tau)$ in a form very similar to that of $\mathcal{P}(t)$ in Eq. (\ref{Pexact}). 
This can be achieved by using $\sigma_n^x(\tau)=e^{\mathcal{H}\tau}\sigma_n^xe^{-\mathcal{H}\tau}$ and inserting the identity $\sum_i|i\rangle\langle i|$ in Eq. (\ref{Cn}), leading ultimately to 
\be 
C_n(\tau)=\sum_{i}|\langle i| \sigma^x_n|0\rangle|^2e^{-\tau(E_i-E_0)}.
\label{Csum}
\ee
Here, $|i\rangle$ denotes the $i$th eigenstate of $\mathcal{H}$ 
with energy $E_i$. 
The excitations of low energy $\epsilon_i=E_i-E_0$ can be identified (with an increasing precision for vanishing $\epsilon_i$) with the spin flips of 
coherent spin clusters produced by the SDRG method, and, 
accordingly, the excitation energy is the twice of the effective transverse-field of the corresponding cluster, $\epsilon_i\simeq 2\tilde h_i$.
If site $n$ is part of a cluster with an effective field $\tilde h_n$ and has  
a weak effective bond $\tilde J_i\ll\tilde h_n$ to a far away cluster with a much smaller excitation energy $\epsilon_i\simeq 2\tilde h_i\ll 2\tilde h_n$, then, by a perturbative calculation, we obtain for the 
matrix element 
$\langle i| \sigma^x_n|0\rangle\simeq \tilde J_i/\tilde h_n$.
The SDRG method provides thus an approximate form of $C_n(\tau)$, which is almost identical to that of $\mathcal{P}(t)$ obtained by the symmetrized method.  
The only difference is that the terms appearing in $C_n(\tau)$ are the {\it squares} of those in $\mathcal{P}(t)$. 
In the critical point, the scaling variable $\gamma=\ln[1/C_n(\tau)]/\ln\tau$ 
will therefore be asymptotically twice the variable $x$ given in Eq. (\ref{gammasum}).
Consequently, the scaling function for the surface spin of a semi-infinite chain is 
\be 
\tilde f_{\rm auto}(\gamma)=\frac{1}{2}e^{-\gamma/2},
\label{auto_1}
\ee
while, for a spin in an infinite chain, it is 
\be 
\tilde f_{\rm auto,b}(\gamma)=\frac{1}{2}\tilde f_{b}(\gamma/2). 
\label{auto_b}
\ee
The rescaled distributions calculated numerically by the free-fermion method in the bulk of the RTIC \cite{ky} and at the surface of the random XX-chain (in which the autocorrelation function is identical to that of the RTIC) \cite{ijr} fit satisfactorily to the functions given in  Eq. (\ref{auto_b}) and Eq. (\ref{auto_1}), respectively, for moderate $\gamma$. The discrepancy with the numerical data for larger $\gamma$ is attributed to the slow convergence discussed in section \ref{num1}.

\section{Summary and outlook}
\label{summary}

We have studied in this paper distributions of dynamical quantities in various models that are tractable by the strong disorder renormalization group method. 
Works on the disordered contact process prior to the present one have 
dealt with the average survival probability.  
Instead, we have considered here the distribution of the survival probability in individual environments, and have shown that the SDRG method is able to describe this quantity. 
According to our results, it shows multi-scaling in the critical point, meaning that, for a fixed, large $t$ it is an inverse power of the time $\mathcal{P}(t)=t^{-\delta}$, but instead of a single value, 
$\delta$ is characterized by a broad distribution in the limit $t\to\infty$. 
By the SDRG method, we have calculated the probability density $\tilde f(\delta)$ in the one-dimensional model analytically in the end point of a semi-infinite lattice, where it is a pure exponential, and numerically in the bulk. In the latter case, we conjecture an algebraic divergence $\sim \delta^{-1+x_b/\psi}$ in the limit $\delta\to 0$ and an exponential decay $\sim e^{-2\delta}$ in the opposite limit $\delta\to\infty$. 
According to the results, universality, i.e. independence from the specific form of the distribution of the parameters in the critical point is predicted by the SDRG method to be valid in a sense wider than known so far. 
It holds, namely, not only for the averages but also for the scaling functions provided by the method, which are conjectured to be exact. 

We have also shown, that certain dynamical quantities of other models described by an IRFP can be calculated similarly to $\mathcal{P}(t)$. 
In the model of random walks in random environments, the survival probability constructed by the SDRG method has been found to have 
the same structure as $\mathcal{P}_1(t)$ of the contact process.  
As a consequence, it shows multi-scaling in the case of a zero mean bias and has the same (exponential) scaling function as $\mathcal{P}_1(t)$ has.
In the random transverse-field Ising chain, we have pointed out that the way of the calculation of the spontaneous magnetization \cite{fisher} is analogous to that of the dominant term of $\mathcal{P}(t)$ at the level of the SDRG and the corresponding scaling functions are therefore identical in the two problems. 
Beyond the dominant term, a more closer similarity between the imaginary-time spin-spin autocorrelation function of the RTIC and $\mathcal{P}(t)$ has been revealed at the level of the SDRG, leading to a simple relationship between the scaling functions in the two problems. 

We have restricted ourselves in this work to one-dimension, but, according to numerical studies, the critical contact process and the RTIC are described by an IRFP even in higher dimensions \cite{kovacs} and in certain networks \cite{jk}, at least for strong enough disorder. 
(As opposed to this, the disorder in the problem of random walks causes only a logarithmic correction to normal diffusion in two dimensions and is irrelevant in higher dimensions \cite{bouchaud}.) 
The method of calculating the dynamical quantities considered here can be extended to and carried out, at least numerically, in higher dimensions, as well. 
Here, the corresponding scaling functions are completely unknown. In particular, it is an open question whether the scaling and non-scaling contributions to the average are comparable in general as it has been found in one-dimension. 
 
We have seen here that drawing random environments from an ensemble, the effective decay exponent characterizing a given environment for a large, fixed $t$ is a random variable with a probability density $\tilde f(\delta)$ in the limit $t\to\infty$. 
However, rather than considering an ensemble, one could take a single (infinite) environment and sample the effective exponents $\delta_n=\ln[1/\mathcal{P}(t_n)]/\ln t_n$ at a sequence of times $t_1,t_2,\dots$, such that $t_n\to\infty$ if $n\to\infty$. 
An interesting question is whether the limit probability density of $\delta_n$ is independent of the underlying environment and, if yes, whether it is identical to $\tilde f(\delta)$. 

A further question, which may be worth considering, is whether dynamical correlations of random quantum spin chains other than that investigated here (and some of which have been studied numerically by the free-fermion technique \cite{ijr}) can be treated by the SDRG method in the way described in the present work.

\acknowledgments
The author thanks discussions with R. Dickman, F. Igl\'oi, I. A. Kov\'acs, and G. \'Odor. 
This work was supported by the J\'anos Bolyai Research Scholarship of the
Hungarian Academy of Sciences, by the National Research Fund 
under grant no. K75324, and partially supported by the European Union and the
European Social Fund through project FuturICT.hu (grant no.:
TAMOP-4.2.2.C-11/1/KONV-2012-0013).

\appendix
\section{Survival probability of mortal random walks}
\label{mortal}

Arranging the probabilities that the walker is on site $i$ at time $t$ 
in a row vector $p(t)=(p_0(t),p_1(t),p_2(t),\dots)$, where $0$ refers to the absorbing site, the rate matrix $W$ of the continuous-time random walk problem described in the text, which is related to the Liouville matrix $Q$ as $W=-Q^T$,  
is given by  
\be
W=
\begin{pmatrix}
0 & 0 & 0& 0&\dots \\
\tilde\mu_1 & -\omega_1 & \tilde\nu_1 & 0 & \dots\\
\tilde\mu_2 & 0 & -\omega_2 &   \tilde\nu_2 & 0\\
\vdots & & & \ddots & \ddots
\end{pmatrix}.
\ee
It has a zero eigenvalue $\epsilon_0=0$ corresponding to the absorbing state 
$(1,0,0,\dots)$ and non-zero eigenvalues 
$\epsilon_n=-\omega_n$, $n=1,2,\cdots$. 
Let us assume that these are non-degenerate, i.e. $\omega_n\neq\omega_m$ if $n\neq m$, which is not much restrictive since, for a continuous distribution of the transition rates of the original model, the degenerate cases are of zero measure.  
The initial state is $p(0)=(0,1,0,0,\dots)$ and writing it as a linear combination of the left eigenvectors $v^{(n)}$ of $W$, 
\be
p(0)=\sum_{n=0}^{\infty}c_nv^{(n)}, 
\label{init}
\ee
the state at time $t$ is given by 
\be 
p(t)=\sum_{n=0}^{\infty}c_nv^{(n)}e^{\epsilon_nt}.
\ee
The survival probability can be written as 
\be 
\mathcal{P}_1(t)=1-p_0(t)=1-\sum_{n=0}^{\infty}c_nv_0^{(n)}e^{\epsilon_nt}=-\sum_{n=1}^{\infty}c_nv_0^{(n)}e^{\epsilon_nt},
\label{cv}
\ee
where we have used that $\lim_{t\to\infty}p_0(t)=c_0v_0^{(0)}=1$. 
Next, let us determine the left eigenvectors $v^{(n)}$ corresponding to non-zero eigenvalues, i.e. $n>0$. 
Introducing the variable $\Delta_n^m\equiv \omega_n+\epsilon_m=\omega_n-\omega_m$
it is easy to show that the components satisfy the recursion relations 
\be 
v_i^{(n)}=\frac{\Delta_{i+1}^{n}}{\tilde\nu_i}v_{i+1}^{(n)}, 
\label{rec}
\ee
by which, together with the relation $\sum_{i=0}^{\infty}v_i^{(n)}=0$ generally valid for non-stationary states (here $n>0$), the eigenvector components can be calculated. Since $\Delta_i^i=0$, one can show that $v_i^{(n)}=0$ for all $n>1$, 
$1\le i\le n-1$. 
Using this, we obtain that the expansion coefficients $c_n$ in Eq. (\ref{init}) satisfy 
\beqn
c_1v_1^{(1)}=1, \nonumber \\
\sum_{i=1}^{(n)}c_iv_n^{(i)}=0,  \qquad n>1.
\eeqn  
Solving these equations recursively results in 
\be 
c_n=-\frac{1}{v_n^{(n)}}\left(\prod_{j=1}^{n-1}\tilde\nu_j\right)
\sum_{i=1}^{n-1}\left(\prod_{j=1,j\neq i}^{n}\Delta_j^i\right)^{-1}.
\label{cn}
\ee
Using the identity 
\be
\sum_{i=1}^{n}\left(\prod_{j=1,j\neq i}^{n}\Delta_j^i\right)^{-1}\equiv 0,
\ee
which can be proved by induction, Eq. (\ref{cn}) simplifies to 
\be 
c_n=\frac{1}{v_n^{(n)}}\prod_{j=1}^{n-1}\frac{\tilde\nu_j}{\Delta_j^{n}}.
\ee
The coefficients in Eq. (\ref{cv}) are thus 
\be 
-c_nv_0^{(n)}=-\frac{v_0^{(n)}}{v_n^{(n)}}\prod_{j=1}^{n-1}\frac{\tilde\nu_j}{\Delta_j^{n}}.
\ee
The ratio $\frac{v_0^{(n)}}{v_n^{(n)}}$ can be calculated by using the recursion relations of components in Eq. (\ref{rec}), resulting in
\be 
\frac{v_0^{(n)}}{v_n^{(n)}}=-s_n,
\ee
where $s_n$ is given in Eq. (\ref{sn}). 

\section{Corrections to the dominant term}
\label{corrections}

First, let us consider the terms in Eq. (\ref{prw0}) having indices $n>M$: 
\be
C_+=\sum_{n=M+1}^{\infty}p_ne^{-\Omega_n/\Omega}. 
\ee
We can write an upper bound for the ratio of $C_+$ and the dominant term $T_M$ 
\be 
\frac{C_+}{T_M}<\frac{\sum_{n=M+1}^{\infty}p_n}{p_Me^{-\Omega_M/\Omega}}
<e\left[\frac{\tilde\nu_M}{\omega_M}+\frac{\tilde\nu_M}{\omega_M}\frac{\tilde\nu_{M+1}}{\omega_{M+1}}+\dots\right],
\ee
since $\Omega_M<\Omega<\Omega_{M-1}$. According to Eq. (\ref{fpsol}),  
$\frac{\tilde\nu_n}{\omega_n}=O[(\Omega_n/\Omega_0)^{\eta_n}]$, where $\eta_n$ is exponentially distributed, therefore  the upper bound in the above inequality tends to zero stochastically in the limit $\Omega_M\to 0$. 

Next, let us consider the leading term $T_{M-1}$ among those with $n<M$. 
If $T_M$ and $T_{M-1}$ were comparable, then $T_{M-1}$ would give a correction of $O(\Gamma^{-1})$ to $\ln[1/\mathcal{P}_1(t)]/\ln t$, which tends to zero in the scaling limit. So the contribution of the term $T_{M-1}$ (and those with $n<M$) is certainly negligible if $T_{M-1}<T_M$ or, equivalently, 
\beqn 
\ln\frac{T_{M-1}}{T_M}=\ln\left[\frac{\omega_{M-1}}{\tilde\nu_{M-1}}
e^{-\frac{\Omega_{M-1}-\Omega_M}{\Omega}}\right] \nonumber \\
\approx \eta_{M-1}\Gamma_{M-1}-\frac{\Omega_{M-1}}{\Omega}<0,
\eeqn
which leads to 
\be 
\Gamma> \Gamma_{M-1}+\ln(\eta_{M-1}\Gamma_{M-1}).
\ee



\begin{thebibliography}{}


\bibitem{wd}
S. Wiseman and E. Domany, Phys. Rev. E {\bf 52}, 3469 (1995).

\bibitem{im}
F. Igl\'oi  and C. Monthus, Phys. Rep. {\bf 412}, 277 (2005).

\bibitem{vojta_rev}
T. Vojta, J. Phys. A {\bf 39}, R143-R205 (2006).

\bibitem{fisher}
D.S. Fisher, Phys. Rev. Lett. {\bf 69}, 534 (1992);
Phys. Rev. B {\bf 51}, 6411 (1995).

\bibitem{fdm}
D. S. Fisher, P. Le Doussal, and C. Monthus, Phys. Rev. Lett. {\bf 80} 3539 (1998);
P. Le Doussal, C. Monthus, and D. S. Fisher, Phys. Rev. E {\bf 59}, 4795 (1999).

\bibitem{sk}
F. Solomon, {\it Ann. Prob.} {\bf 3}, 1 (1975);
H. Kesten, M. V. Kozlov, and F. Spitzer {\it Compos. Math.} {\bf 30}, 145 (1975) 

\bibitem{hiv} J. Hooyberghs, F. Igl\'oi, and C. Vanderzande,
Phys. Rev. Lett. {\bf 90}, 100601 (2003); 
Phys. Rev. E {\bf 69}, 066140 (2004).

\bibitem{noest} A.J. Noest, Phys. Rev. Lett. {\bf 57}, 90 (1986); 
Phys. Rev. {\bf B 38}, 2715 (1988).

\bibitem{cp} T.E. Harris, Ann. Prob., {\bf 2}, 969 (1974).

\bibitem{liggett} T.M. Liggett, {\it Stochastic interacting systems:
  contact, voter, and exclusion processes} (Berlin, Springer, 2005).

\bibitem{md} J. Marro and R. Dickman, {\it Non-equilibrium phase
    transitions in lattice models}, Cambridge Univ. Press, (Cambridge
  1999).

\bibitem{odor}
G. \'Odor, {\it Universality in Nonequilibrium Lattice Systems}, World
Scientific (Singapore, 2008); Rev. Mod. Phys. {\bf 76}, 663 (2004).

\bibitem{hhl} 
M. Henkel, H. Hinrichsen, and S. L\"ubeck, {\it Non-equilibrium
  Phase transitions} (Springer, Berlin 2008).  


\bibitem{vojta} T. Vojta and M. Dickison, Phys. Rev. E {\bf 72}, 036126 (2005);
T. Vojta, A. Farquhar, and J. Mast, Phys. Rev. E {\bf 79}, 011111 (2009);
T. Vojta, Phys. Rev. E {\bf 86}, 051137 (2012).

\bibitem{moreira}
A. G. Moreira and R. Dickman, Phys. Rev. E {\bf 54}, R3090 (1996). 

\bibitem{ky}
J. Kisker and A. P. Young, Phys. Rev. B {\bf 58}, 14397 (1998).

\bibitem{ijr}
F. Igl\'oi, R. Juh\'asz, and H. Rieger, Phys. Rev. B {\bf 61}, 11552 (2000).


\bibitem{juhasz}
R. Juh\'asz, Phys. Rev. E {\bf 87}, 022133 (2013). 

\bibitem{igloi}
F. Igl\'oi, Phys. Rev. B {\bf 65}, 064416 (2002).


\bibitem{griffiths} R.B. Griffiths, Phys. Rev. Lett. {\bf 23}, 17 (1969); 
B.M. McCoy, Phys. Rev. Lett. {\bf 23}, 383 (1969).


\bibitem{hv}
J. Hooyberghs and C. Vanderzande, Phys. Rev. E  {\bf 63}, 041109 (2001).

\bibitem{ascpmc}
R. Juh\'asz, J. Stat. Mech. (2013) P10023.


\bibitem{young}
A. P. Young, Phys. Rev. B {\bf 56}, 11691 (1997).

\bibitem{bouchaud}
J. P. Bouchaud and A. Georges, Phys. Rep. {\bf 195}, 217 (1990).

\bibitem{ir}
F. Igl\'oi and H. Rieger, Phys. Rev. E {\bf 58}, 4238 (1998). 

\bibitem{cd}
A. Comtet and D. S. Dean, J. Phys. A.: Math. Gen. {\bf 31}, 8595 (1998).

\bibitem{jsi}
R. Juh\'asz, L. Santen, and F. Igl\'oi, Phys. Rev. E {\bf 72}, 046129 (2005). 

\bibitem{mg}
C. Monthus and T. Garel, J. Phys. A: Math. Theor. {\bf 41}, 255002 (2008).

\bibitem{dp}
B. Derrida and Y. Pomeau, Phys. Rev. Lett. {\bf 48}, 627 (1982).

\bibitem{mdh}
S.K. Ma, C. Dasgupta, and C.-K. Hu, Phys. Rev. Lett. {\bf 43}, 1434 (1979). 

\bibitem{yr}
A. P. Young and H. Rieger, Phys. Rev. B {\bf 53}, 8486 (1996). 

\bibitem{fy}
D. S. Fisher and A. P. Young, Phys. Rev. B {\bf 58}, 9131 (1998). 

\bibitem{ri}
H. Rieger and F. Igl\'oi, Europhys. Lett. {\bf 39}, 135 (1997). 

\bibitem{kovacs} I.A. Kov\'acs and F. Igl\'oi, Phys. Rev. B {\bf 82},
  054437 (2010); Phys. Rev. B {\bf 83}, 174207 (2011).

\bibitem{jk}
M. A. Mu\~noz, R. Juh\'asz, C. Castellano, and G. \'Odor, 
Phys. Rev. Lett. {\bf 105}, 128701 (2010); 
R. Juh\'asz and I. A. Kov\'acs, J. Stat. Mech. (2013) P06003. 





\end{thebibliography}
\end{document}